\documentclass[11pt]{article}
\usepackage{multibox,amssymb,amsfonts,array}
\usepackage[centertags,intlimits]{amsmath}
\usepackage{graphicx}
\usepackage{hyperref}
\usepackage{pstricks}
\usepackage{latexsym}
\usepackage{epic}
\usepackage{epsfig}

\def\be{\begin{equation}}
\def\ee{\end{equation}}
\def\beq{\begin{eqnarray}}
\def\eeq{\end{eqnarray}}
\def\mathun{\hbox{ 1\hskip -3pt l}}
\def\nn{\nonumber}
\def \RR{{\mathbb{R}}}

\def \NN{{\mathbb{N}}}
\def\sst#1{{\scriptscriptstyle #1}}
\def\so{\sst{[0]}}
\def\pa{\partial}
\def\ie{{\it i.e.~}}
\def\a{\alpha}
\def\b{\beta}
\def\g{\gamma}
\def\d{\delta}

\def\l{\lambda}
\def\La{\Lambda}
\def\G{\Gamma}

\def\m{\mu}
\def\n{\nu}

\def\o{\omega}

\def\S{\Sigma}
\def\t{\tau}
\def\p{\pi}

\def\th{\theta}
\def\ca{{\cal A}}
\def\cB{{\cal B}}

\def\ce{{\cal E}}
\def\cf{{\cal F}}
\def\cg{{\cal G}}
\def\ch{{\cal H}}

\def\ck{{\cal K}}
\def\cl{{\cal L}}
\def\cm{{\cal M}}
\def\cn{{\cal N}}

\def\cp{{\cal P}}

\def\cv{{\cal V}}

\begin{document}

\vspace{3mm}

\begin{center}
{\LARGE \bf Describing general cosmological singularities in Iwasawa variables}\\[7mm]

Thibault Damour and Sophie de Buyl\\[6mm]

{\sl Institut des Hautes Etudes Scientifiques\\
35 route de Chartres, 91440 Bures--sur--Yvette, France}\\[15mm]

\begin{tabular}{p{12cm}}
\hline 
\\ 
\hspace{5mm}{\bf Abstract:}  Belinskii, Khalatnikov, and Lifshitz (BKL) conjectured that the description of the asymptotic behavior of a generic solution of Einstein equations near a spacelike singularity could be drastically simplified by considering that the time derivatives of the metric asymptotically dominate (except at a sequence of instants, in the `chaotic case') over the spatial derivatives. We present a precise formulation of the BKL conjecture (in the chaotic case) that consists of basically three elements: (i) we parametrize the spatial metric $g_{ij}$ by means of \emph{Iwasawa variables }($\beta^a,{\cal N}^a{}_i$); (ii)
we define, at each spatial point, a (chaotic) \emph{asymptotic evolution system }Êmade of ordinary differential equations for the Iwasawa variables; and (iii) we characterize the exact Einstein solutions $\beta, \, {\cal{N}}$ whose asymptotic behavior is described by a solution $\beta_{\sst{[0]}}, \, {\cal{N}}_{\sst{[0]}}$ 
of the previous evolution system by means of a `\emph{generalized Fuchsian system}' for the differenced variables $\bar \beta = \beta - \beta_{\sst{[0]}}$, $\bar {\cal N}Ê= {\cal N} - {\cal N}_{\sst{[0]}}$, and by requiring that $\bar \beta$ and $\bar {\cal N}$ tend to zero on the singularity. We also show that, in spite of the apparently chaotic infinite succession of `Kasner epochs' near the singularity, there exists a well--defined \emph{asymptotic geometrical structure} on the singularity : it is described by a \emph{partially framed flag}. Our treatment encompasses Einstein--matter systems (comprising scalar and $p$--forms), and also shows how the use of Iwasawa variables can simplify the usual (`asymptotically velocity term dominated') description of non--chaotic systems. 
\thispagestyle{empty}
\\ \\
\hline
\end{tabular}
\end{center}

\newpage
\tableofcontents
\newpage
\section{Introduction}

The works \cite{Belinsky:1970ew,BKL2,Belinsky:1982pk} of Belinskii, Khalatnikov and Lifshitz (BKL) proposed a description of the asymptotic behavior of the gravitational field in the vicinity of a spacelike singularity of a $D=4$ spacetime satisfying the vacuum Einstein equations. They also investigated the $D=5$ vacuum Einstein case as well as the $D=4$ spacetime with a massless scalar field \cite{BK}. Finally, they analyzed more general Einstein--matter systems (e.g. Einstein--Yang--Mills) in \cite{Belinsky:1988mc}. Extension of the BKL analysis to higher dimensions was addressed within the context of pure gravity in \cite{Demaret:1985js,Demaret:1986ys}. It is convenient to express the BKL analysis using Hamiltonian methods. This leads to considering the motion of a particle in an auxiliary Lorentzian space submitted to the influence of a linear superposition of exponential potential walls \cite{Belinsky:1970ew, Misner:1969hg, Chitre, Misnerb,melni,Damour:2002et}. This approach allows for a relatively easy generalization of the work of BKL to any spacetime dimension and with any $p$--form field content \cite{Damour:2002et}.
\newline

As argued by BKL, a drastic simplification in the Einstein equations occurs near a spacelike singularity (located at proper time $t=0$) in that the partial differential equations for the metric can be essentially replaced by ordinary differential equations with respect to time. In physical terms this corresponds to an effective decoupling of spatial points $x_1 \neq x_2$ as $t\rightarrow 0$. Depending on the specific theory at hand (spacetime dimension, field content, couplings to the dilatons), the BKL approach leads to expect two possible types of behavior:
\begin{itemize}
\item \emph{`Non--chaotic behavior'} (or Monotonic power law):  the spatial scale factors 
(and the dilaton fields $e^\phi$ if any),  behave at each spatial point in a monotone, power-law fashion 
in terms of the proper time as one approaches the singularity at $t=0$, \ie at each spatial point the asymptotic form of the metric looks like a Kasner metric. On the other hand, the 
$p$--form fields $A$ have limits as $t\rightarrow 0$. Theories exhibiting this asymptotic behavior are, for instance, pure gravity in $D\geq 11$ \cite{Demaret:1985js,Demaret:1986ys,Damour:2002tc}, and gravity coupled to a scalar field in any dimensions \cite{BK,Andersson:2000cv}.
\item \emph{`Chaotic behavior'}: at each spatial point, the asymptotic behavior is given by a chaotic succession of an infinite 
number of increasingly shorter Kasner regimes as one goes to the singularity. Important examples of theories exhibiting this asymptotic behavior are pure gravity in $D\leq 10$ \cite{Demaret:1985js,Demaret:1986ys}, and the bosonic sector of all supergravities associated with the low energy limit of string or M-theory \cite{Damour:2000wm}.  
\end{itemize}
The non--chaotic case has been formulated in rigorous mathematical terms by considering an auxiliary asymptotic dynamics called the `asymptotically velocity term dominated' (AVTD) system \cite{Eardley:1971nj}.
The AVTD system is obtained by neglecting all the spatial derivatives in the considered Einstein--matter system. Einstein equations then reduce to ordinary differential equations (ODEs). The solutions of this asymptotic system are precisely given by Kasner--like metrics. Fuchsian methods \cite{Kichenassamy:1997bt} can then be used to prove that, given a solution of the velocity dominated system, there exists a (geometrically unique) solution of Einstein's equations that asymptotically approaches this solution. These Fuchsian methods have been used to mathematically describe cosmological singularities in various simplified contexts: Gowdy spacetimes \cite{Kichenassamy:1997bt}, plane symmetric spacetimes with a massless scalar field \cite{Rendall}, polarized and half-polarized $U(1)$ symmetric vacuum spacetimes \cite{Isenberg:2002jg}, spacetimes with collisionless matter and spherical, plane or hyperbolic symmetry \cite{Rein}, and a particular subset of general Gowdy spacetimes \cite{Chrusciel}. It has also been possible to use Fuchsian methods to mathematically describe singularities without any symmetries: notably for the Einstein--scalar system \cite{Andersson:2000cv}, and for many Einstein--matter models including pure gravity in $D \geq 11$ dimensions \cite{Damour:2002tc}.\newline

By contrast, the general inhomogeneous chaotic case has not yet been tackled by rigorous mathematical methods. The BKL conjectural behavior has been consolidated by recent Iwasawa--variable based analytical treatments \cite{Damour:2002et} and is also supported by numerous numerical results \cite{WIB,U1,bgimw,BIW,garf,Garfinkle:2007rv}. However there exist neither a clear general formulation of the precise asymptotic behavior advocated in the BKL approach, nor any mathematical theorems concerning its compatibility with Einstein field equations. \emph{The purpose of this paper is to present a mathematically precise formulation of the BKL conjecture in the chaotic case. }In other words, we aim at providing a chaotic analog of the AVTD formulation of the non--chaotic case. More precisely, we shall describe the asymptotic dynamics of the gravitational field for Einstein--matter systems, at each spatial point, by a well defined \emph{asymptotic evolution system made of ODEs }. \newline

\begin{figure}[h]
\begin{picture}(0,0)%
\includegraphics{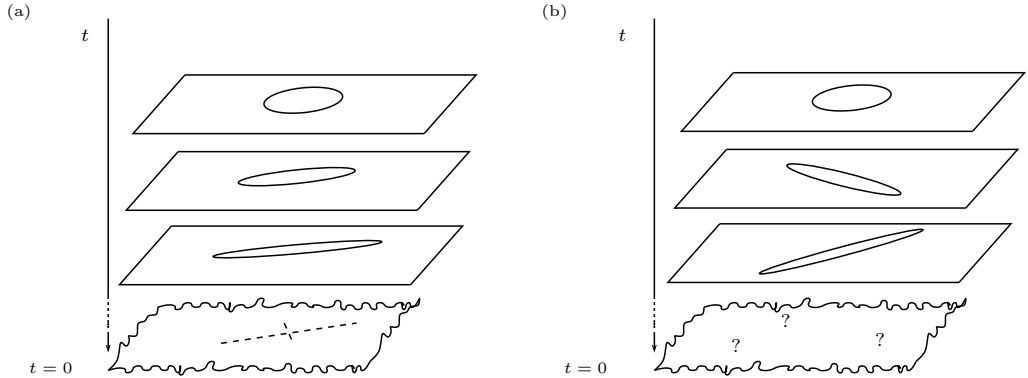}%
\end{picture}%
\setlength{\unitlength}{1243sp}%
\begingroup\makeatletter\ifx\SetFigFont\undefined%
\gdef\SetFigFont#1#2#3#4#5{%
  \reset@font\fontsize{#1}{#2pt}%
  \fontfamily{#3}\fontseries{#4}\fontshape{#5}%
  \selectfont}%
\fi\endgroup%
\begin{picture}(20328,7416)(361,-6799)
\put(14806,-6316){\makebox(0,0)[lb]{\smash{\SetFigFont{7}{8.4}{\familydefault}{\mddefault}{\updefault}{\color[rgb]{0,0,0}?}%
}}}
\put(17686,-6181){\makebox(0,0)[lb]{\smash{\SetFigFont{7}{8.4}{\familydefault}{\mddefault}{\updefault}{\color[rgb]{0,0,0}?}%
}}}
\put(15796,-5821){\makebox(0,0)[lb]{\smash{\SetFigFont{7}{8.4}{\familydefault}{\mddefault}{\updefault}{\color[rgb]{0,0,0}?}%
}}}
\put(811,-6721){\makebox(0,0)[lb]{\smash{\SetFigFont{6}{7.2}{\familydefault}{\mddefault}{\updefault}{\color[rgb]{0,0,0}$t=0$}%
}}}
\put(11476,-6721){\makebox(0,0)[lb]{\smash{\SetFigFont{6}{7.2}{\familydefault}{\mddefault}{\updefault}{\color[rgb]{0,0,0}$t=0$}%
}}}
\put(361,389){\makebox(0,0)[lb]{\smash{\SetFigFont{6}{7.2}{\familydefault}{\mddefault}{\updefault}{\color[rgb]{0,0,0}(a)}%
}}}
\put(1846,-106){\makebox(0,0)[lb]{\smash{\SetFigFont{7}{8.4}{\familydefault}{\mddefault}{\updefault}{\color[rgb]{0,0,0}$t$}%
}}}
\put(11026,389){\makebox(0,0)[lb]{\smash{\SetFigFont{6}{7.2}{\familydefault}{\mddefault}{\updefault}{\color[rgb]{0,0,0}(b)}%
}}}
\put(12556,-106){\makebox(0,0)[lb]{\smash{\SetFigFont{7}{8.4}{\familydefault}{\mddefault}{\updefault}{\color[rgb]{0,0,0}$t$}%
}}}
\end{picture}
\caption{\label{kasnerdirect} \small (a) \emph{Non--chaotic behavior }: sufficiently close to the singularity, the dynamics of the gravitational field can be approximated by a Kasner--like  metric at each spatial point. Let us focus on one particular spatial point where asymptotically the metric is given by $ds_{spatial}^2 =  (t^{2p_1} l_i l_j + t^{2p_2} m_i m_j +.. .+ t^{2p_d} r_i r_j) dx^i \, dx^j$. When $t \rightarrow 0$, the directions for which the Kasner exponent $p_i$ is negative are stretched while the ones with positive exponent are squeezed. At the singularity, these directions are still defined. (b) \emph{Chaotic behavior }: now instead of a Kasner--like  metric at each spatial point, there is a never ending chaotic succession of Kasner epochs before reaching the singularity. Are there still some preferred directions at the singularity? or is some other structure of the metric preserved asymptotically?  }
\end{figure}

In addition to formulating a precise conjecture for the chaotic BKL behavior, we also address the question of whether or not geometrical structures  can be defined \emph{at} the singularity and what are these asymptotic geometrical structures. 
According to the billiard picture, most of the metric variables possess well defined limits at the singularity: the `\emph{off--diagonal variables}', \ie all the variables except the diagonal metric components (and the dilaton)\footnote{A short review of the billiard picture is presented in section \ref{toda}.}. This means in particular that for these variables, initial data can be assigned \emph{at} the singularity. The other variables, \ie the \emph{diagonal variables, } have no limit at the singularity.
Although the `off--diagonal' variables have finite limits, they are (co)frame (and gauge) dependent and thereby they do not have a priori a clear geometrical meaning. Nevertheless, we can wonder whether it is possible to extract some geometrical information from these asymptotic values. 
It turns out that this is possible, but that this \emph{asymptotic geometrical structure} is less `rigid' in the chaotic case at hand than it was in the non--chaotic case. In the non--chaotic case, the asymptotic geometrical structure is simple to describe. The solution is asymptotically given, at each spatial point, by a Kasner--like metric  \cite{Andersson:2000cv,Damour:2002tc}. The (spatial) Kasner metric is, in $d$ spatial dimensions, 
\begin{equation}
g_{ij} (t) = t^{2p_1} l_i l_j + t^{2p_2} m_i m_j +.. .+ t^{2p_d} r_i r_j \, ,\label{kasner}
\end{equation}
where the $p_i$'s ($i=1,2,3$) are the Kasner exponents subject to the Kasner conditions. [Note that this metric possesses a curvature singularity at $t=0$ and that the distances are no longer defined at this singularity, since  either $g_{ij} \underset{t \rightarrow 0}{\rightarrow} \infty$ or $g_{ij} \underset{t \rightarrow 0}{\rightarrow} 0$.] The \emph{Kasner coframes}, \ie the coframes that diagonalise, at each spatial point, the second fundamental form $k_{ij}$ with respect to $g_{ij}$ have finite limits at the singularity (up to independent rescalings they are simply given by $\o^1_K = l_i dx^i, \, \o^2_K = m_i dx^i, \, ... , \o^d_K = r_i dx^i$) and therefore provide a basis of preferred directions, \ie a `directional frame' (and co--frame).  See Figure \ref{kasnerdirect}.\newline

For chaotic systems, we could have expected, from the BKL description of the asymptotic dynamics of the metric as a never ending chaotic succession of Kasner epochs at each spatial point, that no privileged directions can be defined at the singularity especially in view of the effect, discovered in \cite{BKL2}, of a `rotation' of Kasner frames between two successive Kasner epochs.  However, we show in section \ref{pff} that an asymptotic geometrical structure can be defined at the singularity. This structure is less precise than a frame but more precise than a \emph{flag} and therefore we will call it a \emph{partially framed flag}. The precise meaning of this notion is explained in the sequel.  \newline

This paper is organised as follows. We first review in section \ref{toda} the Iwasawa--variables `cosmological billiards' of \cite{Damour:2002et} in order to introduce our notation and stress important features for our purposes. Then, to gain some intuition for how to define the asymptotic system of evolution equations parametrizing a `generic' solution of an Einstein--matter system in the chaotic case, we revisit in section \ref{nonchaos} the non--chaotic case treated in \cite{Andersson:2000cv,Damour:2002tc}. Our new approach is based on Hamiltonian methods and Iwasawa variables, which simplify the previously done analyses.\footnote{A comparison between the two analyses is done in appendix \ref{appB}.} More precisely, we define an asymptotic system of evolution equations and we rewrite the Hamiltonian Einstein--matter evolution equations in terms of the difference between the solution of the full evolution equations and the solution of this asymptotic system. Then we argue that the so--obtained `differenced system' of equations is of the Fuchsian type\footnote{We recall Fuchs' theorem in appendix \ref{appFuchs}.}.  We also treat the constraints by defining asymptotic constraints which, when they are satisfied, imply the vanishing of the exact constraints.  
Next we turn to our main purpose in section \ref{chaos}, that is to give a mathematically precise formulation of the chaotic BKL behavior. This is achieved by defining again an asymptotic system of evolution equations which is a system of ordinary differential equations though it is not necessarily an AVTD system. We then formally rewrite the Hamilton equations in terms of the difference between the solution of the full evolution equations and the solution of the asymptotic system. Finally, we argue that a stronger version of the usual Fuchs theorem is likely to remain valid in the chaotic case. In the last section, we show that for chaotic systems, \emph{partially framed flags} are the asymptotic geometrical structures that stay well defined at the singularity.\newline

We mention the recent paper \cite{Heinzle:2007kv} that establishes the relationship between the Iwasawa--based billiard approach used here and the \emph{dynamical systems approach} to cosmological singularities.


\section{Appearance of Toda--like walls in Einstein--matter Hamiltonians in  Iwasawa variables \label{toda}}

The general systems considered are of the following form
\beq &&S[ g_{\m\n}, \phi, B^{\sst{(p)}}] = \int d^D x \, \sqrt{- 
^{\sst{(D)}}g} \;
\Bigg[R (g) - \partial_\m \phi \partial^\m \phi \nn \\
&& \hspace{2.5cm} - \frac{1}{2} \sum_p \frac{1}{(p+1)!} e^{\l_p
\phi} F^{\sst{(p)}}_{\m_1 \cdots \m_{p+1}} F^{\sst{(p)}  \, \m_1 \cdots \m_{p+1}}
\Bigg] + \dots .~~~~~~~ \label{keyaction} \eeq 
Units are
chosen such that $16 \pi G_N = 1$,  $G_N$ is Newton's
constant and the spacetime dimension $D \equiv d+1$ is left
unspecified. Besides the standard Einstein--Hilbert term the above
Lagrangian contains a dilaton 
field $\phi$ and a number of $p$--form fields $B^{\sst{(p)}}_{\m_1 \cdots \m_p}$ (for 
$p\geq 0$).  The generalization to any number of dilatons is
straightforward. The $p$--form field strengths $F^{\sst{(p)}} = dB^{\sst{(p)}}$ are
normalised as \beq F^{\sst{(p)}}_{\m_1 \cdots \m_{p+1}} = (p+1)
\partial_{[\m_1} B^{\sst{(p)}}_{\m_2 \cdots \m_{p+1}]} \equiv
\partial_{\m_1} B^{\sst{(p)}}_{\m_2 \cdots \m_{p+1}} \pm p \hbox{
permutations }. 
\nn
\eeq As a
convenient common formulation we have adopted the Einstein conformal
frame and normalised the kinetic term of the dilaton $\phi$ with
weight one with respect to the Ricci scalar. The Einstein metric
$ g_{\m\n}$ has Lorentz signature $(- + \cdots +)$ and is used
to lower or raise the indices; its determinant is denoted by $
^{\sst{(D)}}g$. The dots in the action (\ref{keyaction}) above
indicate possible modifications of the field strength by
additional Yang--Mills or Chapline--Manton-type couplings
\cite{Bergshoeff:1981um,Chapline:1982ww}. 
The real parameter $\l_p$ measures the strength
of the coupling of $B^{\sst{(p)}}$ to the dilaton. In the following, for simplicity, we shall treat the case where there is no dilaton $\phi$ and indicate what changes occur when $\phi$ is present.

\subsection{Iwasawa variables \label{iwasawasection}}

Let us give a schematic review of the Iwasawa--variable cosmological billiards. For a detailed derivation, we refer to \cite{Damour:2002et}.  \newline

We choose a slicing of the spacetime we want to construct, $\cm_D = M_d \times \RR$, such that the singularity occurs at the coordinate  time $\tau=+\infty$. We shall define the time slicing $\tau$ by requiring that the `rescaled lapse' $\tilde N = N / \sqrt{g}$ (where $g=$det$g_{ij}$) is equal to some given (weight --1) time--independent density $\mu_{-1}(x)$ on $M_d$. For simplicity, we take $\m_{-1}(x)=1$ in the coframe $\o^i$ we use, so that $\tilde N = 1 \iff N=\sqrt{g}$. In other words our time coordinate is linked to the `proper time' $dt =-Nd\tau$ by $d\tau = - dt / \sqrt{g}$. The slicing is built by use of pseudo--Gaussian coordinates defined by a vanishing shift $N^i=0$, lapse $N(\tau,x^i) = \sqrt{g(\tau,x^i)} \mu_{-1}(x^i) =\sqrt{g(\tau,x^i)} $ and metric
\beq
ds^2 =-(N(\tau,x^i) d\tau)^2 + g_{{ij}}(\tau,x^i) \o^{i} (x^k) \o^{i}(x^k) \, . \label{metric}
\eeq
Here $\o^i(x)= \o^{i}{}_j (x)dx^j$ is a coframe on the given (analytic) spatial manifold $M_d$.\footnote{Note that in \cite{Damour:2002et}, a coordinate basis is used instead of a general basis $\o^i$ but the generalization is straightforward as long as $\o^i = \o^{i}{}_j(x^k) dx^j$ does \emph{not} depend on time.} 
One of the useful technical tools we shall employ here consists in replacing the  $d(d+1)/2$ metric variables $g_{ij}$,  by a new set of variables: $d$ `diagonal degrees of freedom' $\b^a$, together with $d(d-1)/2$ `off--diagonal degrees of freedom' $\cn^a{}_{i} $ where $\cn$ is restricted to be an upper triangular  matrix ($\cn^a{}_{i} =0,$ if $i<a$) with ones on the diagonal ($\cn^a{}_{i} = 1,$ if $ a=i$), such that 
\beq 
g_{ij}= \sum_{a=1}^d e^{-2 \b^a} \cn^a{}_{i} \cn^a{}_{j} \, . \label{iwa}
\eeq
We shall refer to the algebraic decomposition (\ref{iwa}) as the \emph{Iwasawa decomposition} of the metric\footnote{Indeed, it is linked to the Iwasawa decomposition of the vielbeins $\cv^a{}_{i} \in SL(d,\RR)$ (such that $g_{ij}= \cv^a{}_{i}\cv^a{}_{j}$) which reads $\cv = KAN$ where $K \in SO(d)$, $A$ is a diagonal matrix and $N$ is a nilpotent matrix. The Iwasawa variables are uniquely specified by requiring that $K \in SO(d)$ be the unit matrix.}. In $d=3$, the components of the metric read explicitly,
\begin{eqnarray}
g_{11} &=& e^{-2 \beta^1}, \; \; \; g_{12} = \cn^1{}_2 e^{-2 \beta^1},
\; \; \;
g_{13} = \cn^1{}_3 e^{-2 \beta^1}, \nn \\
g_{22} &=& (\cn^1{}_2)^2 e^{-2 \beta^1} + e^{-2 \beta^2}, \; \; \;
g_{23} = \cn^1{}_2 \cn^1{}_3 e^{-2 \beta^1} + \cn^2{}_3 e^{-2 \beta^2}, \nn \\
g_{33}  &=& (\cn^1{}_3)^2 e^{-2 \beta^1} + (\cn^2{}_3)^2 e^{-2 \beta^2} +
e^{-2 \beta^3}
\end{eqnarray}
from which one gets (uniquely)
\begin{eqnarray}
\beta^1 &=& - \frac {1}{2} \ln g_{11}, \; \;  \; \beta^2 = - \frac
{1}{2} \ln \left[ \frac{g_{11} g_{22} -g_{12}^2} {g_{11}} \right] ,  \nn \\
\beta^3 &=& - \frac {1}{2} \ln \left[ \frac {g} {g_{11} g_{22}
-g_{12}^2}\right], \; \;  \; \cn^1{}_2 = \frac {g_{12}}{g_{11}}, \nn \\
\cn^1{}_3 &=& \frac {g_{13}}{g_{11}}, \; \;  \; \cn^2{}_3 = \frac {g_{23}
g_{11} -g_{12} g_{13}}{g_{11} g_{22} -g_{12}^2}.
\label{iwaexplicit} 
\end{eqnarray}
In the cosmological context, one could refer to the `diagonal metric variables' $e^{- \b^a}$ as the `scale factors'. In \cite{Damour:2002et} 
the $\b$'s and the dilaton (when present) were collectively denoted 
$ \b^\mu = \{ \b^a, \, \phi\}$. In the case considered here (no dilaton), we shall use labels from the beginning of the latin alphabet $(a,b,c,..,e)$ to denote the `diagonal variables' $\b^a$. All other variables are called `off--diagonal variables' and are denoted $Q$, 
\beq
Q = \{ \cn, \, \cB^{\sst{(p)}} \} \, , \nn  
\eeq
where $\cB^{\sst{(p)}}$ are the $B^{\sst{(p)}}$ expressed in the generalized Iwasawa coframe $\th^a_{\sst{\mathrm{iwa}}}$, $\th^a_{\sst{\mathrm{iwa}}} := \cn^a{}_{{j}} \o^{{j}} $, {\it e.g.} we have $B_{i_1... i_p}=: \cn^{a_1}{}_{i_1} ... \cn^{a_p}{}_{i_p} \cB_{a_1...a_p}$. Note that, in the Iwasawa coframe, the metric is diagonal: $g^{\sst{\mathrm{iwa}}}_{ab} = e^{-2 \b^a} \d_{ab}$.


\subsection{Hamiltonian approach in Iwasawa variables}

The Hamiltonian action corresponding to the action (\ref{keyaction}) in any pseudo-Gaussian gauge, and in the temporal gauge for the form fields ($B_{0i_1...i_{p-1}} = 0$), reads\footnote{The term $ \pi_\phi \dot{\phi}$ should be added (inside the parenthesis) in the action if a dilaton is present. }
\beq
&& S\left[ g_{ij}, \pi^{{ij}}, B^{(p)}_{{j_1 \cdots  j_p}},
\pi_{(p)}^{{j_1 \cdots j_p}}\right] = \nonumber \\
&& \hspace{1cm}
\int dx^0 \int d^d x \left( \pi^{{ij}} \dot{g}_{{ij}}
+ \frac{1}{p!}\sum_p \pi_{(p)}^{{{j_1 \cdots j_p}}}
\dot{B}^{(p)}_{{j_1 \cdots j_p}} - H \right)
\label{GaussAction}
\eeq
where the Hamiltonian density\footnote{Note that $H$ is a density of weight 1 while $\ch$ is a density of weight 2.} $H$ is\footnote{If a dilaton is present, the term $ \frac{1}{4} \pi_\phi^2 $ should be added in $\ck$ as well as exponential coupling $e^{ \lambda_p \phi}$ in front of the term 
$g \, F^{(p)}_{j_1 \cdots j_{p+1}} F^{(p) \, j_1 \cdots j_{p+1}}$ in $\cm$ and 
$e^{- \lambda_p \phi}$ in front of 
$ \pi_{(p)}^{{j_1 \cdots j_p}} \pi_{(p) \, {j_1 \cdots j_p}}$ in $\ck$.%
The term $g g^{{ij}} \partial_{{i}} \phi \partial_{{j}} \phi$ should also be added to $\cm$.
}
\beq\label{Ham}
H &\equiv&  \tilde{N} \ch \\
\ch &=& \pi^{{ij}}\pi_{{ij}} - \frac{1}{d-1} \pi^i_{\;i} \pi^j_{\;j}
 + \sum_p  \frac{1}{2 \, p!} \, \pi_{(p)}^{{j_1 \cdots j_p}} \pi_{(p) \, {j_1 \cdots j_p}} \nn \\
&-& g R  + \sum_p  \, g \frac{1}{2 \; (p+1)!}\, F^{(p)}_{j_1 \cdots j_{p+1}} F^{(p) \, j_1 \cdots j_{p+1}}
\eeq
where $R$ is the spatial curvature scalar. The dynamical
equations of motion are obtained by varying the above action
w.r.t. the spatial metric components, the spatial
$p$--form components and their conjugate momenta. In addition,
there are constraints on the dynamical variables,
\beq
\ch &\approx& 0  \; \; \; \; \hbox{(``Hamiltonian constraint")},  \label{hamconst}\\
\ch_i &\approx& 0  \; \; \; \; \hbox{(``momentum constraint")}, \label{momconst}\\
\varphi_{(p)}^{{j_1 \cdots j_{p-1}}} &\approx& 0 \; \; \; \;
\hbox{(``Gauss law" for each $p$-form) } \label{Gauss}
\eeq
with\footnote{If there are dilatons, the term $
\pi_{\phi}\pa_i\phi$ should be added to $\ch_i$} 
\beq
\ch_i &:=& -2 {\pi^{j}}_{{i}|{j}} + \sum_p \frac1{p!} \
\pi_{(p)}^{{j_1 \cdots j_p}} F^{(p)}_{{i j_1 \cdots j_{p}}} \\
\varphi_{(p)}^{{j_1 \cdots j_{p-1}}} &:=&
{\pi_{(p)}^{{j_1 \cdots j_{p-1} j_p}}}_{\vert {j_p}}
\eeq
where the subscript $|{j}$ stands for the spatially covariant derivative.\newline

As shown in \cite{Damour:2002et} (and as we will see explicitly for some of the terms below) the Hamiltonian density  of weight 2, $ {\cal{H}}$, expressed in the Iwasawa variables has the following structure:  
\beq {\cal{H}} [\beta,Q; \pi,P] =  \ck + \cv \, ,  \label{general_ham}   \eeq
where
\beq
\ck &=& \frac{1}{4}G^{ab} \pi_a \pi_b \nn \\
\cv &=&  \sum_{A}c_A(Q,P,\partial_x \b, \partial^2_x \beta, \partial Q, \partial^2 Q) e^{-2 w_{A}(\b)} \nn \, . \eeq 
Here $G^{ab}$ is the inverse of the quadratic form $G_{ab}$ which is defined by $G_{ab}d\b^a d\b^b:= \sum_{a=1}^d (d\b^a)^2 -(\sum_{a=1}^d d\b^a)^2$. Note the important fact that this metric has a Lorentzian signature $(-,+,...,+)$. $P$ stands for $\{ \cp^{{i}}{}_a, \, {\cal{E}}^{a_1...a_p}_{(p)} \}$, where the $\cp^{{i}}{}_a$'s are the momentum conjugate to the $\cn^a{}_{{i}}$'s and the ${\cal E}^{a_1...a_p}_{(p)}$'s are the $\pi^{{i_1...i_p}}_{(p)}$ --- \ie the momentum conjugate to the $B^{\sst{(p)}}_{{i_1...i_p}}$'s --- expressed in the Iwasawa frame. Note that the $\cp^i{}_a$'s are strictly lower diagonal (they exist only when $i>a$ and vanish when $i\leq a$). \newline
 
Note the special structure of the (weight 2) Hamiltonian density ${\cal{H}}$ with  (i) a kinetic term $\pi^2$ for a `point particle' of coordinates $\b^a$ moving in (ii) a sum of `exponential walls' (for their $\b$ dependence). We shall refer to them as \emph{Toda walls} $e^{ -2w_A(\b)}$ where the $w_A(\b)$ are certain linear forms in the $\b$'s. `Toda' refers to the well known Toda models involving such exponential walls. For instance, the kinetic terms of the off--diagonal degrees of freedom $\cn$ in $H$ give, when expressed in Iwasawa variables, terms proportional to $e^{-2(\b^b -\b^a)}$ for $b >a$ with coefficient proportional to $\cp^2$. Therefore the kinetic terms for the $\cn$'s furnish the walls $w_{S \, ab} = \b^b - \b^a$ called `symmetry walls'. The kinetic terms of the $p$--forms in $H$ yield a sum of terms proportional to $e^{-2w_e(\b)}$ where $w_{e \, a_1...a_p} (\b) = \b^{a_1}+...+ \b^{a_p} $ (`electric $p$--form walls'). The curvature term $-g R$ in $H$ gives terms proportional to $e^{-2 w_{abc}(\b)}$ where $w_{abc}(\b)=  \b^a + \sum_{e \neq b, c (b \neq c)} \b^e$ (`curvature walls') and their coefficients, when $a \neq b$, $a \neq c$ and $b \neq c$, are given by $(C_{\mathrm{\sst{iwa}}}{}^a{}_{bc})^2$. Here, the $C_{\mathrm{\sst{iwa}}}$ are the structure functions of the Iwasawa coframe $\th^a_{\mathrm{\sst{iwa}}}=  \cn^a{}_j \th^j $, $d\th^a_{\mathrm{\sst{iwa}}} = -{1\over 2} \, C_{\mathrm{\sst{iwa}}}{}^a{}_{bc} \th^b_{\mathrm{\sst{iwa}}} \wedge \th^c_{\mathrm{\sst{iwa}}}$. Note that the structure functions $C_{\mathrm{\sst{iwa}}}$ depend on the $\cn$'s and the $\partial_x \cn$'s. \newline

A heuristic analysis of the BKL limit indicates the crucial role played by the linear forms $w_A(\b)$. Indeed, in this limit the walls become infinitely sharp and are located at the hyperplanes given by the linear forms $w_A(\b) = 0$, the motion is then restricted to the region of $\b$--space defined by the inequalities $\{ w_A(\b) \geq 0\}$. The set of \emph{dominant walls} is the minimal subset $\{w_{\ca}(\b)\}$ --- the indices $\ca$ belong to a subset of the indices $A$ --- such that the subset of inequalities $w_{\ca}(\b) \geq 0$ implies the full set of inequalities $w_{A}(\b) \geq 0 \,  \forall A$.\footnote{The dominant linear forms can be identified in many physically relevant cases with the simple roots of an hyperbolic Kac--Moody algebra \cite{Damour:2000hv,Damour:2001sa,Damour:2002fz}.} A crucial consistency condition for these definitions, which is found to be satisfied for all models, is that the coefficients of the dominant walls be positive: $c_{\ca} \geq 0$.We decompose the set of indices $\{ A\}$ for the walls into the set of indices for the dominant walls $\{ \ca \}$ and the remaining ones  (`subdominant walls') $\{ \ca^\prime \}$. Moreover, another crucial structure of the potential is that the dependence of the wall coefficients on spatial derivatives in such that, 
\beq
\cv =   \sum_{\sst{\ca}}c_{\sst{\ca}}(Q,P, \partial_x Q) e^{-2 w_{\sst{\ca}}(\b)} 
+ \sum_{ \sst{\ca}^\prime }c_{\sst{\ca}^\prime}(Q,P,\partial_x \b, \partial^2_x \beta, \partial Q, \partial^2 Q) e^{-2 w_{\sst{\ca}^\prime}(\b)}
\label{potentiel} \, ,
\eeq
where the coefficients of the dominant walls are found \emph{never} to depend on $\{ \partial_x \b, \partial^2_x \beta  \}$. \newline

For instance, in the case of pure gravity in $d$ space dimensions, the $d$ dominant walls comprise
\begin{itemize}
\item[(i)] $d-1$ dominant `symmetry' walls $w_{S \, a-1 \, a}(\b)= \b^a -\b^{a-1}$ ($a =2,...,d$) and
\item[(ii)] one curvature wall $w_{1\, d-1 \, d} (\b) = 2\b^1  + \b^2 + ...+ \b^{d-2}$. Note that in $d=3$, $w_{123}=2 \b^1$ corresponds to the (first of the) famous BKL walls of the form $a^4 + b^4 +c^4$, where $a = e^{-2 \b^1},\,b = e^{-2 \b^2},\,c = e^{-2 \b^3}$ \cite{Belinsky:1970ew}.
\end{itemize} 


\subsection{Hamilton evolution equations}
 
Let us indicate, in a sketchy manner, the structure of the Hamilton evolution equations following from (\ref{general_ham}), 
\beq 
\partial_\tau \beta^a &=& {1 \over 2}G^{ab}\pi_b \nn \, , \\ 
\partial_\tau \pi_a &=& \sum_A \left( 2 c_A w_{A\,a} e^{-2 w_A(\b)} + \partial_x ({\partial c_A \over \partial \partial_x \beta^a} e^{-2 w_A(\beta)}) - \partial^2_x ({\partial c_A \over \partial \partial^2_x \beta^a} e^{-2w_A(\beta)}) \right) \nn \, , \\ 
\partial_\tau Q &=& \sum_A {\partial c_A \over \partial P} e^{-2w_A} \nn \, ,  \\
\partial_\tau P &=& \sum_A \left(-{\partial c_A \over \partial Q} e^{-2w_A} + 
\partial_x({\partial c_A \over \partial \partial_x Q} e^{-2w_A})-
\partial^2_x({\partial c_A \over \partial \partial^2_x Q} e^{-2w_A}) \right)\, , \label{general_eq}
\eeq
where  $c_A=c_A(Q,P,\partial_x \b, \partial^2_x \beta, \partial Q, \partial^2 Q) $, $w_{A\, a}$ denotes the (covariant) components of the linear forms $ w_A(\beta) =w_{A\, a}\beta^a $. The system (\ref{general_eq}) is the one that we will analyze in detail in the sequel. \newline

Let us recall the basic classification of the set of dominant walls: either the fundamental chamber defined by the dominant inequalities $w_\ca (\b) \geq 0$ is contained within the future ($\sum_{a=1}^d \b^a > 0$) light cone $G_{ab} \b^a \b^b=0$, or it is not. The first case define what we call here \emph{chaotic systems}, while the second defines \emph{non--chaotic systems}. For instance, pure gravity in $D=d+1$ is chaotic for $d\leq 9$ , and non--chaotic for $d>9$. Note that this classification does \emph{not} correspond to the often used asymptotically--velocity--terms--dominated (AVDT) systems versus non--AVTD ones. Indeed, there are AVTD systems that are chaotic. For instance, the Einstein--Maxwell system (in any dimension $D=d+1$) is always chaotic and we shall see below that its asymptotic chaotic behavior can be described as a naive AVTD truncation of the full dynamics. Let us also mention that they are non--AVTD chaotic systems that are equivalent to AVTD chaotic system. For instance, gravity coupled to a $(d-2)$--form is chaotic and non--AVTD because driven by its magnetic wall. However, by Hodge duality, it is equivalent to the Einstein--Maxwell system which is chaotic and AVDT. \newline

As a warm up towards understanding the structure of the evolution equations (\ref{general_eq}) we shall first consider the so--called non--chaotic systems.


\section{Iwasawa--variables treatment of non--chaotic systems \label{nonchaos}}

In this section, we reformulate the results of \cite{Andersson:2000cv,Damour:2002tc} by using the Iwasawa variables, within an Hamiltonian approach. Let us recall that the treatment used in \cite{Andersson:2000cv,Damour:2002tc} consisted of rewriting Einstein--matter systems into a \emph{Fuchsian} form, \ie  
\beq
\pa_\tau u - \ca u = e^{-\mu \tau} f (x,\tau,u,\pa_x u) \, , \label{fuchstau}
\eeq
where $\m >0$ and where the crucial conditions are (i) that the source term $f$ should be bounded when $\tau \rightarrow \infty$ (while the other variables take their values in a bounded set) and (ii) that the eigenvalues of the (space and time independent) matrix $\ca$ be strictly larger than $-\m$ \cite{choquet-bruhat}; see appendix \ref{appFuchs} for precise mathematical conditions. Then the main result of the Fuchs theorem is that there exists a unique solution $u(\tau,x)$ of (\ref{fuchstau}) which tends to zero as $\tau \rightarrow 0$. Moreover, the exponential decay of the source $ e^{-\m \tau}$ imposes a corresponding fast decay of solution which we shall write as
\beq
u = O (e^{-\m^{(-)} \tau}) \, , \nn
\eeq
where $\mu^{(-)}$ can be any number satisfying $0<\mu^{(-)}<\m$ [Note that $\mu^{(-)}$ can be as close as we want to $\m$]. \newline

Here we are going to show that the evolution equations in Iwasawa variables given by equations (\ref{general_eq}) can be rewritten in an alternative Fuchsian form which leads to a streamlined derivation of the results of \cite{Andersson:2000cv,Damour:2002tc}. In order to do that, we need to do two things (i) define an asymptotic evolution system whose solutions $\{\b_{\sst{[0]}},\pi_{\sst{[0]}},Q_{\sst{[0]}},$ $P_{\sst{[0]}}\}$ parametrize the generic asymptotic exact solutions $\{\b,\pi,Q,P\}$, (ii) rewrite the system of equations (\ref{general_eq}) in terms of the differences  $u$ between $\{\b,\pi,Q,P\}$  and $\{\b_{\sst{[0]}},\pi_{\sst{[0]}},Q_{\sst{[0]}},$ $P_{\sst{[0]}}\}$ such that the system of equations for $u$ is Fuchsian, and (iii) define asymptotic constraints in such a way that the exact constraints are satisfied if the asymptotic constraints and the asymptotic equations of motion are fulfilled. This is done in the sequel and implies by the Fuchs theorem that there is a unique solution $u$ that vanishes when $\tau$ goes to infinity; the Fuchs theorem also tells us how $u$ goes to zero as $\tau \rightarrow \infty$. This result gives a precise sense in which the approximate solutions $\{\b_{\sst{[0]}},\pi_{\sst{[0]}},Q_{\sst{[0]}},$ $P_{\sst{[0]}}\}$ parametrize the asymptotic behavior of the exact solutions $\{\b,\pi,Q,P\}$. If the asymptotic solutions  $\{\b_{\sst{[0]}},\pi_{\sst{[0]}},Q_{\sst{[0]}},P_{\sst{[0]}}\}$ are general enough, it means that we have found the general asymptotic behavior of the gravitational field in the vicinity of a spacelike singularity. [More precisely, we want here that the solutions of the asymptotic system together with the associated asymptotic constraints contain the same number of arbitrary functions which is expected to enter the general solution of the exact constrained Einstein--matter system.] We will see that our new formulation is significantly simpler than that of \cite{Andersson:2000cv,Damour:2002tc} and is suggestive for approaching of the chaotic case. \newline


\subsection{Definition of the asymptotic evolution equations}

The first step is to define a simplified system of equations that describe the asymptotic dynamics of the fields. There are, a priori, several choices for defining an asymptotic system when using Iwasawa variables. For instance, we can either neglect certain terms directly in the Hamiltonian or neglect some terms in the equations of motion. One of these choices gives a system essentially equivalent to the usually considered AVTD system in \cite{Andersson:2000cv,Damour:2002tc}. It would consist in keeping \emph{only the symmetry walls} in the Hamiltonian (\ref{general_ham}). Here, we will consider a technically simpler choice consisting in neglecting \emph{all the walls}. Concretely, this means that we \emph{define }the `asymptotic Hamiltonian' as
\beq \ch_\circ[\beta_{\sst{[0]}},Q_{\sst{[0]}}; \pi_{\sst{[0]}},P_{\sst{[0]}}] = {1\over 4} G^{ab} \pi_{\sst{[0]} \, a} \pi_{\sst{[0]}\, b} \label{asym_ham} \, , \eeq 
the $\sst{[(0)]}$'s refer to the zeroth order approximation of our general solution. The Hamilton equations corresponding to the Hamiltonian (\ref{asym_ham}) are 
\beq
\partial_\tau \b_{\sst{[0]}}^a &=&  {1\over 2} G^{ab}\pi_{\sst{[0]} \,_b} \nn \, , \\
\partial_\tau \pi_{\sst{[0]} \, a} &=& 0 \nn \, , \\
\partial_\tau Q_{\sst{[0]}} &=& 0 \nn \, , \\
\partial_\tau P_{\sst{[0]}} &=& 0 \, . \label{asym_eq}
\eeq
The solutions of these Hamilton equations are schematically (suppressing indices),
\beq
\beta_{\sst{[0]}}&=& p_\circ \tau + \b_\circ \, , \nn \\
\pi_{\sst{[0]}}&=& p_\circ \, ,\nn \\
Q_{\sst{[0]}} &=& Q_\circ \, , \nn \\
P_{\sst{[0]}} &=& P_\circ \, ,\label{asym_sol}
\eeq where $p_\circ, \, \b_\circ, \, Q_\circ, \, P_\circ$ do not depend on the time but depend on the spatial coordinates $x^i$.\footnote{As is usual when discussing Fuchsian theorems one makes the technical assumption that the spatial dependence of all the initial data $(p_\circ(x),\b_\circ(x),...)$ is real analytic.} Note that the metric corresponding to the `asymptotic solution' (\ref{asym_sol}) \emph{does not} generically corresponds to a Kasner--type metric (\ie a metric of the type (\ref{kasner})). Indeed, the Iwasawa `off--diagonal' variables $\cn$'s of a generic Kasner metric have limits as $\tau \rightarrow \infty$ but are $\tau$ dependent for finite $\tau$, while the $\cn^a{}_{{i}} $'s corresponding to the solution (\ref{asym_sol}) are constants [see paragraph 4.2 of \cite{Damour:2002et} for explicit expression of the Iwasawa variables of a Kasner metric]. 


\subsection{Definition of the asymptotic constraints}

As the asymptotic Hamiltonian constraint, it is natural to take the asymptotic Hamiltonian (\ref{asym_ham}), 
\beq
\ch_{\sst{[0]}} =  {1\over 4} G^{ab} \pi_{\sst{[0]} \, a} \pi_{\sst{[0]} \, b}  \label{Casym} \, , \eeq
which has the useful property of being conserved along the asymptotic evolution equations (\ref{asym_eq}). 
Concerning the asymptotic momentum constraints, we need to know their structure in Iwasawa variables to be able to conclude. In view of this, we first express the momentum conjugate to the metric $g_{{ij}}$ in terms of Iwasawa variables.  \newline

Let $\pi_{\sst{\mathrm{iwa}}}{}^{ab} := \cn^a{}_{{i}}\cn^b{}_{{j}} \pi^{{ij}}$ denote the Iwasawa--frame components of the momentum conjugate to the $g_{{ij}}$. Using this definition, the effect of the transformation of the configuration variables $\{ g_{{ij}} \} \rightarrow \{\b^a,\cn^a{}_{{i}} \}$  on their conjugate momenta is obtained from writing
\beq
\dot g_{{ij}}  \pi^{{ij}}&=& 
\sum_a 2e^{-2 \beta^a}( \dot \cn^a_{{i}} - \dot \b^a \cn^a_{{i}}) \cn^{-1 \, {i}}{}_c  \pi_{\sst{\mathrm{iwa}}}{}^{ca} \nn \\
&=& \dot \b^a \pi_a + \dot \cn^a_{{i}} \cp^{{i}}{}_a \, , \nn 
\eeq 
from which we can extract that (we recall that the metric in the Iwasawa--frame is $g^{\sst{\mathrm{iwa}}}_{ab} = e^{-2\b^a} \d_{ab}$ so that $\pi_{\sst{\mathrm{iwa}}}{}^a{}_b = e^{-2 \b^b} \pi_{\sst{\mathrm{iwa}}}{}^{ab}$)  
\beq 
\pi_{\sst{\mathrm{iwa}}}{}^b{}_b  &=& -\frac{1}{2} \p_b \hspace{.8cm} \mathrm{no \,  \,  sum \, \, over \,  } b ,\nn \\
 \cn^{-1 \, {i}}{}_c \pi_{\sst{\mathrm{iwa}}}{}^c{}_a &=& \frac{1}{2} \cp^{{i}}{}_a \, \hspace{.8cm} only \, \,\mathrm{for \,  \,   } {i} > a. \nn
 \eeq 
In order to invert the above formula and get the $\pi_{\sst{\mathrm{iwa}}}{}^c{}_a$ in terms of the $\pi_a$ and $\cp^{{i}}{}_a $, let us rewrite the above equation for all ${i}$ and $ a$ as follows,
\beq
\cn^{-1 \, {i}}{}_{c\, (+)} \pi_{\sst{\mathrm{iwa}}}{}^c{}_a = \frac{1}{2} \cp^{{i}}{}_{a\, [-]}  + X_{(+)} \, , \label{inversion1}
\eeq
where $X$ is a matrix defined by this equation, and where we have added to the various triangular matrices that appear an index referring to the fact that it is an upper/lower triangular matrix $(+)/(-)$ or a strictly upper/lower triangular matrix $[+]/[-]$. We can now multiply equation (\ref{inversion1}) by $\cn = \cn_{(+)}$ and obtain,
\beq
\pi_{\sst{\mathrm{iwa}}}{}^b{}_a = \frac{1}{2} \cn^b{}_{{i} \, (+)}\cp^{{i}}{}_{a\, [-]}  + \cn^b{}_{{i} \, (+)}X_{(+)} \, . \label{inversion}
\eeq
Let us decompose the matrix 
$\pi_{\sst{\mathrm{iwa}}}{}^b{}_a$ into its strictly lower triangular part 
$\pi_{\sst{\mathrm{iwa}}}{}^b{}_{a\, \sst{[-]}}$, its diagonal part 
$\pi_{\sst{\mathrm{iwa}}}{}^b{}_b  = -\frac{1}{2} \p_b$ and its strictly upper triangular part 
$\pi_{\sst{\mathrm{iwa}}}{}^b{}_{a\, \sst{ [+]}}$. The projection of both sides of equation (\ref{inversion}) on their strictly lower triangular parts yields an explicit expression for $ \pi^b{}_{a \, \sst{\mathrm{iwa}[-]}}$ (for $b>a$), namely   
$
\pi^b{}_{a \, \sst{\mathrm{iwa}[-]}}  = \frac{1}{2} \cn^b{}_{{i} \, (+)}\cp^{{i}}{}_{a\, [-]}  \th(b-a)  , $ where 
\beq
\th (x) := \left\{ \begin{array}{ll}
                      0  & \mbox{if $x\leq0$} \\
                      1 & \mbox{if $x >0$.}
                      \end{array}
                      \right. 
\nn 
\eeq
Note now that   
$\pi_{\sst{\mathrm{iwa}}}{}^b{}_{a\, \sst{ [+]}}$, being obtained from the \emph{symmetric} matrix $\pi_{\sst{\mathrm{iwa}}}{}^{ab} = \pi_{\sst{\mathrm{iwa}}}{}^{ba}$ by lowering an index by the metric  $g_{\sst{\mathrm{iwa}} \, ab} = e^{- 2 \b^a} \d_{ab}$, can be related to 
$\pi_{\sst{\mathrm{iwa}}}{}^b{}_{a\, \sst{ [-]}} $ in the following way,
\beq
\pi_{\sst{\mathrm{iwa}}}{}^b{}_{a\, \sst{[+]}} = e^{-2(\b^a - \b^b)} 
\pi_{\sst{\mathrm{iwa}}}{}^a{}_{b\, \sst{ [-]}}  \, . \label{pilower}
\eeq 
Finally, we have the following links, 
\beq
\pi_{\sst{\mathrm{iwa}}}{}^b{}_b  &=& -\frac{1}{2} \p_b \hspace{.2cm} \mathrm{no \,  \,  sum \, \, over \,  } b \, ,\nn \\
\mathrm{if \, b> a}\hspace{.8cm} \pi_{\sst{\mathrm{iwa}}}{}^b{}_a &=& \pi_{\sst{\mathrm{iwa}}}{}^b{}_{a \, \sst{[-]}}  = \frac{1}{2} \cn^b{}_{{i}}\cp^{{i}}{}_{a} \,  , \nn \\
\mathrm{if \, a> b}\hspace{.8cm} \pi_{\sst{\mathrm{iwa}}}{}^b{}_a&=&\pi_{\sst{\mathrm{iwa}}}{}^b{}_{a \, \sst{[+]}}   =\frac{1}{2}  e^{-2(\b^a - \b^b)}\cn^a{}_{{i}}\cp^{{i}}{}_{b} \,  , \label{pivsp}
\eeq
Therefore, the $\pi_{\sst{\mathrm{iwa}}}{}^a{}_{b\, \sst{[-]}} $ are linear in $ \cp^{{i}}{}_a $ and $\cn^{{i}}{}_a$, while the $\pi_{\sst{\mathrm{iwa}}}{}^a{}_{b \, \sst{[+]}} $'s depend also on the $\b$'s though the $e^{-2(\b^a - \b^b)}$ with $a>b$ (see equation (\ref{pilower})), \ie through \emph{symmetry walls} $e^{-2w_{S \, ba}(\b)}$.\newline

Let us now express the momentum constraints in Iwasawa variables $\ch_a$. They read,\footnote{The comma in the expression $g_{cd,b}$ denotes the spatial derivative in the Iwasawa frame, \ie $g_{cd,b} = e_{\sst{\mathrm{iwa}\, }}{}_b{}^i e_i ( g_{cd})$ where $e_{\sst{\mathrm{iwa} }}{}_b =e_{\sst{\mathrm{iwa} }}{}_b{}^i e_i$ is the Iwasawa frame (dual to the coframe $\th^b_{\sst{\mathrm{iwa} }}$: $\th_{\sst{\mathrm{iwa} }}{}^b (e_{\sst{\mathrm{iwa} }}{}_a) = \d^b_a$)
and where $e_i = e_i{}^j \pa_j$ is the frame dual to the basic coframe $\o^i  = \o^i{}_j dx^j$ used in equation (\ref{metric}). We will also sometimes denote $g_{cd,b}$  by $\pa_b g_{cd}$.}
  \beq
 -{1\over 2}\ch_a &=& \nabla_b \pi_{\sst{\mathrm{iwa}}}{}^b{}_a = \partial_b \pi_{\sst{\mathrm{iwa}}}{}^b{}_a + 
 \Gamma^b_{db}\pi_{\sst{\mathrm{iwa}}}{}^d{}_a  
   - \Gamma^d_{ab} \pi_{\sst{\mathrm{iwa}}}{}^b{}_d 
- \frac{1}{2} g^{cd}g_{cd,b} \pi^b{}_a \, ,\nn   \\
& & + {1 \over p!} \ce^{a_1...a_p} \cf^{\sst{(p)}}_{a a_1...a_p}\nn \, ,
\eeq
where the $\cf^{\sst{(p)}}_{a a_1...a_p}$'s are the $F^{\sst{(p)}}_{i i_1...i_p}$'s expressed in the Iwasawa basis and the $\G^a{}_{bc}$'s are the connection coefficients (with $c$ denoting the differentiation index) in the Iwasawa basis, 
\beq 
\G^a{}_{bc} &=& {1\over 2} e^{2 \b^a}( \d_{ab}e^{-2 \b^b},_c +\d_{ac}e^{-2 \b^c},_b-\d_{bc}e^{-2 \b^b},_a) \nn \\
 & & + {1\over 2}(- C_{\sst{\mathrm{iwa}}}{}^a{}_{bc} + e^{-2 (\b^b-\b^a)}C_{\sst{\mathrm{iwa}}}{}^b{}_{ac}+ e^{-2 (\b^c-\b^a)}C_{\sst{\mathrm{iwa}}}{}^c{}_{ab}) \, . \nn
\eeq
Note that $\pi_{\sst{\mathrm{iwa}}}{}^a{}_b$ is a tensorial density of weight 1. The $C_{\sst{\mathrm{iwa}}}{}$ are the structure functions of the coframes $\th_{\sst{\mathrm{iwa}}}^a$, they are related to the structure functions $C$ in the coframe $\o^{{i}}$ by the formula,
\beq
-{1 \over 2} C_{\sst{\mathrm{iwa}}}{}^a{}_{bc} = \pa_c \cn^a{}_{{i}}\cn^{-1\, {i}}{}_b -{1 \over 2}  \cn^a{}_{{i}} C^{{i}}{}_{{jk}}\cn^{-1\, {j}}{}_b \cn^{-1\, {k}}{}_c \label{iwastructurefunctions}\eeq
Inserting these results in the expression for the momentum constraints gives
\beq -{1\over 2}\ch_a &=& \partial_b \pi^b{}_a + C_{\sst{\mathrm{iwa}}}{}^c{}_{cb} \pi^b{}_a + C_{\sst{\mathrm{iwa}}}{}^d{}_{ac} \pi^c{}_d - {1 \over 2} (\pa_a \b^d) \pi_d \nn \\
& & + {1 \over p!} \ce^{a_1...a_p} \cf^{\sst{(p)}}_{a a_1...a_p} \, .\label{momconstgeneral}\eeq
Note that this is the general expression for the momentum constraints expressed in the Iwasawa variables. \newline

We then \emph{define }the asymptotic momentum constraints by discarding the $\pi_{\sst{\mathrm{iwa}}}{}^b{}_{a \, \sst{[+]}}$ contributions in the exact constraint (\ref{momconstgeneral}): 
\beq
-{1 \over 2}\ch_{a \sst{[0]}} &:=& \big[\pa_b \pi_{\sst{\mathrm{iwa}}}{}^b{}_{a \, [-]} -{1 \over 2} \pa_a\pi_a+ C_{\sst{\mathrm{iwa}}}{}^c{}_{cb} \pi_{\sst{\mathrm{iwa}}}{}^b{}_{a\, [-]} + C_{\sst{\mathrm{iwa}}}{}^d{}_{ac} \pi_{\sst{\mathrm{iwa}}}{}^c{}_{d \, [-]} \nn \\
& &-{1 \over 2}C_{\sst{\mathrm{iwa}}}{}^c{}_{ca} \pi_a -{1 \over 2} C_{\sst{\mathrm{iwa}}}{}^d{}_{ad} \pi_d
- {1 \over 2} ( \b^d_{,a}) \pi_d \nn \\
& & + {1 \over p!} \ce^{a_1...a_p} \cf^{\sst{(p)}}_{a a_1...a_p} \big]^{[0]}\, , \hspace{1cm} \mathrm{no \, sum \, over \, a, \, sum \, over \, d}\label{Caasym} 
\eeq
where the overall bracket $[ \, \, ]^{[0]}$ means that one must do the following replacements $\pi \rightarrow \pi_{\sst{[0]}}, \, \b \rightarrow \b_{\sst{[0]}}, \, Q \rightarrow Q_{\sst{[0]}}, \, P \rightarrow P_{\sst{[0]}}$. 
Finally this definition corresponds (besides the replacement $\{ \b, \pi,Q,P\} \rightarrow \{ \b_{\sst{[0]}}, \pi_{\sst{[0]}},Q_{\sst{[0]}},P_{\sst{[0]}}\}$) to setting to zero the symmetry walls $e^{-2(\b^a-\b^b)}$ (with $a>b$) in the full momentum constraints.
Along the solution (\ref{asym_sol}) of the evolution equations (\ref{asym_eq}), the only time dependent term in $\ch_{a \sst{[0]}}$ is $-   \pa_a\b^d_{\sst{[0]}} \pi_{ \sst{[0]}\, d} / 2 = - \tau  \, \pa_a p_\circ^d \pi_{\circ \, d} /2$ so that we have the following relation, 
\beq \partial_\tau \ch_{a \sst{[0]}} = \pa_a \ch_{\sst{[0]}} \hspace{2cm} \mathrm{modulo \, equations} \, (\ref{asym_eq}). \eeq 
From this relation, we conclude that, when the Hamiltonian $\ch_{\sst{[0]}}=0$ constraint is satisfied, the momentum constraints are conserved when the asymptotic evolution system (\ref{asym_eq}) is satisfied. Finally, it suffices to impose the constraints $\ch_{\sst{[0]}}$ and $\ch_{a \sst{[0]}}$ at any fixed moment to guarantee that they are satisfied for all time. \newline

Similarly, the asymptotic Gauss constraint for each $p$--form is defined to be the Gauss constraint with the asymptotic variables $Q_{\sst{[0]}}, \, P_{\sst{[0]}} $ instead of $Q, \, P$:
\beq 
\varphi^{a_1... a_{p-1}}_{\sst{(p) \, [0]}} &: =& \pa_{a_p} \pi_\so{}^{a_1...a_p}
-{1\over 2} C_{\sst{\mathrm{iwa}} \, \so}{}^{a_1}{}_{b a_p}  \pi_\so{}^{ba_2...a_p} -... - 
{1\over 2} C_{\sst{\mathrm{iwa}} \, \so}{}^{a_{p-1}}{}_{b a_p}  \pi_\so{}^{a_1...ba_p} \nn \\
& &+ \,  C_{\sst{\mathrm{iwa}} \, \so}{}^{a_p}{}_{a_p b} \pi_\so{}^{a_1...a_{p-1}b} \, .
\label{asympgaussconst}
\eeq
These constraints are preserved by the time evolution since the $\pi_\so{}^{a_1...a_p}$ (which are  some of the $P_{\sst{[0]}} $'s) and the $C_{\sst{\mathrm{iwa}} \, \so}{}^{a}{}_{b c}$ (which depend on the $\cn_\so \in Q_{\sst{[0]}}$'s via 
(\ref{iwastructurefunctions})),  are constants according to the asymptotic evolution equations (\ref{asym_eq}). 

 
\subsection{Construction of a Fuchsian system for the `differenced variables'}
 
Let us introduce the differences $\bar \b, \, \bar \pi, \, \bar Q, \, \bar P$ via\footnote{where here $ \beta_{\sst{[0]}}, \, \pi_{\sst{[0]}}, \, Q_{\sst{[0]}}$ and $P_{\sst{[0]}}$ are given in (\ref{asym_sol}).}
\beq\beta  &=& \beta_{\sst{[0]}} + \bar \beta \, , \nn \\
\pi  &=& \pi_{\sst{[0]}} +\bar \pi \, , \nn \\
Q  &=& Q_{\sst{[0]}} + \bar Q\, , \nn \\
P &=& P_{\sst{[0]}} + \bar P\, , 
\label{fullsol}\eeq 
and express the Hamilton equations in term of these variables. 
This gives
\beq \partial_\tau  \bar \beta - {1\over 2}\bar \pi &=& 0 \nn \\
\partial_\tau  \bar \pi &=& 
2 c_{A} w_{A} e^{-2w_{A}(\beta_{\sst{[0]}})} e^{- w_A(\bar \beta)}\nn \\ &+&
\partial_x ( { \partial c_{A} \over \partial \partial_x \b} e^{-2w_{A} (\beta_{\sst{[0]}} ) }e^{-2 w_A(\bar \beta)} ) \nn  \\
&-&
\partial^2_x ( { \partial c_{A} \over \partial \partial^2_x \b} e^{-2w_{A} (\beta_{\sst{[0]}} ) }e^{- 2 w_A(\bar \beta)} )
\nn \\
\partial_\tau  \bar Q &=& {\partial c_A \over \partial P} e^{-2w_{A}(\beta_{\sst{[0]}})} e^{- w_A(\bar \beta)} \nn \\
\partial_\tau \bar P &=& -{\partial c_A \over \partial Q} e^{-2w_{A}(\beta_{\sst{[0]}})} e^{- w_A(\bar \beta)} + 
\partial_x({\partial c_A \over \partial \partial_x Q} e^{-2w_{A}(\beta_{\sst{[0]}})} e^{- w_A(\bar \beta)}) \nn \\
& &-
\partial^2_x({\partial c_A \over \partial \partial^2_x Q} e^{-2w_{A}(\beta_{\sst{[0]}})} e^{- w_A(\bar \beta)})\label{fuchsnonchaotic} \eeq
where $c_A=c_A(Q_{\sst{[0]}}+\bar Q,P_{\sst{[0]}}+ \bar P,\partial_x (\b_{\sst{[0]}} + \bar \beta ), \partial^2_x  (\b_{\sst{[0]}}+ \bar \beta ) ,\partial (Q_{\sst{[0]}}+\bar Q), \partial^2 (Q_{\sst{[0]}}+\bar Q )) $.
Let us sketch the proof that this system will be Fuchsian if all the `walls' $w_A(\b)$ entering the equation (\ref{fuchsnonchaotic}) are such that the following conditions hold, 
\beq
\forall A ,\, \forall x \in U, \, \, w_{A}(p_\circ(x)) 
>\epsilon > 0 \, ,  \label{condinit}
\eeq
where the $p_\circ$ is the initial datum entering equation (\ref{asym_sol}) which must also satisfy the constraint (\ref{Casym}), \ie $G^{ab} p_{ \circ \, a}p_{\circ \, b} =0$, as well as the asymptotic momentum constraints (\ref{Caasym}). Here, $U$ denotes some open domain within the analytic $d$--dimensional manifold, on which one applies the Fuchs theorem. 
The fact that the system (\ref{fuchsnonchaotic}) is indeed of the form (\ref {fuchstau}) for $u = (\bar \b, \bar \pi, \bar Q, \bar P)$ comes from two separated facts. First the matrix $\ca$ being
\beq 
\left(
\begin{array}{cccc}
0 & {1\over 2} & 0 & 0 \\
0 & 0 & 0 & 0 \\
0 & 0 & 0 & 0 \\
0 & 0 & 0 & 0 
\end{array}
\right) \, , \label{matrixA}
\eeq 
is a nilpotent matrix and therefore is (thanks to the recent progress concerning Fuchsian systems \cite{choquet-bruhat}) an allowed matrix $\ca$ for a Fuchs system (see appendix \ref{appFuchs}). 
Concerning the source term, let us show why the conditions (\ref{condinit}) guarantee that the `source term $f$' --- \ie the right hand side of the system of equations (\ref{fuchsnonchaotic}) --- satisfies the right properties. Essentially the Fuchsian conditions boil down to requiring that the source term should be of order $O(e^{-\mu \tau})$ for some $\mu >0$ when $\{\bar \b,\bar \pi, \bar Q, \bar P,\pa_x \bar \b,\pa_x \bar \pi, \pa_x \bar Q,\pa_x \bar P, \pa^2_x \bar \b, \pa^2_x \bar Q\}$ take their values in a bounded set while $\tau \in [\tau_\circ,+\infty]$ (see appendix \ref{appFuchs}). 
The explicit time dependence of the source has three origins
\begin{itemize}
\item[1.] the Toda walls whose ($\tau$) time dependence is exponential $e^{-2w_A(\b_{\sst{[0]}})}=e^{-2w_A(p_{\circ})\tau -2 w_A(\b_\circ)}$
\item[2.] the various space derivatives appearing in the r.h.s. of (\ref{fuchsnonchaotic}) can `bring down', when operating on $e^{-2w_A(p_{\circ}(x))\tau} $, one ($\pa_x$) or two ($\pa^2_x$) powers of $\tau$.
\item[2.] In addition, the dependence of the wall coefficients $c_A$ on $\pa_x \b_{\sst{[0]}}$ and  $\pa^2_x \b_{\sst{[0]}}$ means that, for some walls\footnote{The walls depending on spatial derivatives of $\b$ are only `subdominant' gravitational walls, see \cite{Damour:2002et}.}, the coefficient $c_A$ can also involve one or two powers of $\tau$.
\end{itemize}
Summarising, the r.h.s. of (\ref{fuchsnonchaotic}) is a sum of terms of the form 
$$ P(\tau) e^{-2w_A(p_\circ) \tau}\, ,$$
where $P(\tau)$ is a \emph{polynomial} in $\tau$. If we choose a $\m$ \emph{strictly smaller } than all the quantities $2w_A(p_\circ(x)) > 2\epsilon >0$ considered for any $x \in U$ and any type of wall $A$, we can conclude that  all the source terms in the equations (\ref{fuchsnonchaotic}) are of the required order $O(e^{-\m \tau})$ for $\tau \in [\tau_\circ, \infty]$ and $x \in U$. Therefore we can conclude that there exists a unique solution $\{ \bar \b(\tau,x),\bar \pi(\tau,x), \bar Q(\tau,x), \bar P(\tau,x)\}$ of equations (\ref{fuchsnonchaotic}) that vanishes when $\tau \rightarrow \infty$. Moreover, this unique solution satisfies (within the considered spatial domain $U$) the following estimate as $\tau \rightarrow + \infty$: 
\beq 
\bar \b &=& O(e^{-\m^{(-)}\tau}) \,, \nn \\
\bar \pi &=& O(e^{-\m^{(-)}\tau}) \,, \nn \\
\bar Q &=& O(e^{-\m^{(-)}\tau}) \,, \nn \\
\bar P &=& O(e^{-\m^{(-)}\tau}) \,, \nn 
\eeq
where $0<\mu^{(-)}<\mu$ ($\mu$ being strictly smaller than the quantities $2w_A(p_\circ)>2\epsilon$).


\subsection{Constraints}

It remains to show that if the asymptotic equations of motions, and the asymptotic constraints  (\ref{Casym}, \ref{Caasym}, \ref {asympgaussconst}) are satisfied, then the exact constraints (\ref{hamconst}, \ref{momconst}, \ref{Gauss}) will also be satisfied. \newline

Let us first deal with the Gauss constraints (for notational simplicity, we shall consider the case of one $p=1$--form, \ie a Maxwell field).
We recall, from equation (\ref{Gauss}) that, in this case, the exact Gauss constraint (in Iwasawa variables) reads, 
\beq 
\varphi:= \nabla_a \pi^a = \pa_a \pi^a +  C_{\sst{\mathrm{iwa}}}{}^b{}_{ba}Ê\pi^a \approx 0 \, . \label{exactgauss1}
\eeq
This exact constraint is preserved by the exact equations of motions. 
We also recall that the asymptotic Gauss constraint (\ref {asympgaussconst}), in our case, read, 
\beq 
\varphi_\so:= \nabla_a \pi_\so{}^a = \pa_a \pi_\so{}^a + C_{\sst{\mathrm{iwa}}}{}_\so{}^b{}_{ba}Ê\pi_\so{}^a \approx 0\, , \label{gauss1}
\eeq
and is conserved modulo the asymptotic equations of motions (\ref{asym_eq}). We impose that the asymptotic Gauss constraint (\ref{gauss1}) hold. 
Then, the fact that the difference between the exact and asymptotic Gauss constraints are given by exponential `walls' (entering the differences between the $\pi$ and $\pi_\so$ etc) implies that the exact constraint (\ref{exactgauss1}) vanishes when $\tau \rightarrow + \infty$. Finally, from the obvious fact that an asymptotically vanishing quantity which is constant must be zero, we can conclude that the exact Gauss constraints are (weakly) fulfilled. In particular, the relationships $\nabla_\m T^{\m \n}=0$ are satisfied. \newline

Let us now consider the Hamiltonian and momentum constraints. To show that they are satisfied, we will argue that their evolution system is `Fuchsian' in some generalized sense and therefore that there will be a \emph{unique} solution of this system that vanishes when $\tau \rightarrow \infty$. The Fuchsian system in question turns out to be homogeneous, so that the unique solution that vanishes when $\tau \rightarrow +\infty$ \emph{must be} exactly zero.  Then we will discuss why the exact constraints vanish when $\tau \rightarrow +\infty$ (when the asymptotic constraints are satisfied) and conclude that, since they do, they must be the unique vanishing solution, \ie zero. \newline

Let us first write down the evolution system satisfied by the exact constraints as a consequence of the Bianchi identities and the exact evolution equations. This reads,\footnote{See equations (\ref{evoeqchgf}) and  (\ref{eveqchigf}) in  appendix \ref{hamform}. These equations were derived in a coordinate basis and without matter. It is obvious how to get the expression in a general basis. Moreover, we do not have to consider matter since $\nabla_\m T^\m{}_\n=0$ [as a consequence of the Gauss constraints (which are imposed to be satisfied) and the matter equations of motion]. Therefore, we can just replace in equations (\ref{evoeqchgf}) and (\ref{eveqchigf}) in appendix \ref{hamform} the expressions (\ref{ch}) and (\ref{chi}) by their general expressions (\ref{hamconst}) and (\ref{momconst}).
}
\beq
 \partial_\tau \ch &=& Êe^{-2 \sum_b \b^b}(\nabla^a \ch_a -2 \sum_c \b^c_{,a} \ch^a) \, , \nn \\
 \partial_\tau \ch_a - \nabla_a\ch &=& 0
\, , \label{evoexconst}
\eeq
where the covariant spatial derivatives must take into account the weights of the various densities: $\ch_a$ has weight 1, and $\ch$ has weight 2.  The `source term', \ie the r.h.s. of the evolution equation for $\ch$,  can be rewritten as, 
\beq 
\sum_a e^{-2 \mu_a(\b)} (\pa_a \ch_a + 2 \b^a_{\, , a} \ch_a -2(\sum_c \b^c)_{,a} \ch_a)\, ,
\nn 
\eeq
where $\mu_a(\b) = \sum_{b\neq a} \b^b $ is a subdominant curvature wall (it corresponds to the special case of the curvature walls $w_{abc}(\b)$ with $a=b$). Therefore, the source term is an allowed one for the Fuchsian--like system (\ref{fuchstau}) with $u=(\ch,\ch_a)$ and $\mu$ strictly smaller than all the quantities $2\mu_a(p_\circ(x)) > \tilde \epsilon >0$ considered for any $x \in U$. 
However, this homogeneous system is not really a Fuchsian system because of the presence of the spatial derivatives term $ \nabla_a \ch$ in the second equation. The references \cite{Andersson:2000cv,Damour:2002tc} tackled this problem\footnote{We translate here the argument used in \cite{Andersson:2000cv,Damour:2002tc} in our choice of  variables and gauge.} essentially by working with a suitably redefined $\ch$ constraint, say $\bar \ch = e^{\eta \tau} \ch$. This redefinition produces for $\{ \bar \ch,\ch_a \}$ a Fuchsian system if $0<\eta<2 \mu_a(p_\circ)$ (for all $a$ and for all $x \in U$, $p_\circ(x)$ being as in equation (\ref{asym_sol})). However, we think that there might be other ways of dealing with this problem. First, we expect that a generalization of the Fuchs theorem exists for linear homogeneous systems of the type (\ref{evoexconst}), stating that the unique solution that vanishes when $\tau \rightarrow + \infty$ is everywhere zero. We then expect that one way to prove such a theorem is to work with an extended set of variables $\{\ch,\ch_a,\cg_a =\pa_a \ch\}$ and use the recent reference \cite{choquet-bruhat}. To summarise, the system (\ref{evoexconst}) is a Fuchsian--like system that possess a unique solution (which is zero because the system is homogeneous) that vanishes when $\tau \rightarrow + \infty$. 
 \newline

Moreover (i) as the definition of the asymptotic constraints $\ch_{\sst{[0]}}(\b_{\sst{[0]}}, \pi_{\sst{[0]}},Q_{\sst{[0]}},P_{\sst{[0]}})$ and $\ch_{a \, \sst{[0]}}(\b_{\sst{[0]}}, \pi_{\sst{[0]}},Q_{\sst{[0]}},P_{\sst{[0]}})$ differs from the exact ones by neglecting some explicit exponential walls in their mathematical expression,  (ii) as  $\b-\b_{[0]}$ etc tends to zero (essentially as $O(e^{-\m \tau})$) and finally (iii) as we have imposed the asymptotic constraints, we can conclude that the exact constraints $\ch(\b,\p,Q,P)$ and $\ch_a(\b,\p,Q,P)$ tend to zero as $\tau \rightarrow + \infty$. \newline

Finally, the constraints $\ch$ and $\ch_a$ being uniquely defined as being an asymptotically vanishing solution of an \emph{homogeneous} Fuchsian system, vanish for all times $\tau$.
\newline

\begin{figure}[h]
 \caption{ \label{kasnerlikecase} \small{ \emph{Non--chaotic behavior } This picture is a schematic drawing of the asymptotic dynamics of the `diagonal variables' at a given spatial point $x$, this dynamics is represented in the $\b$--space. The `dashed' arrow represents the asymptotic solution $\b_{\sst{[0]}}$ (which is valid after the last collision on a wall and corresponds to a free motion of the particle $\b$). The exact solution is sketched as a continuous curve. The idea is that the approximate solution $\b_{\sst{[0]}}$ becomes better and better as $\tau \rightarrow + \infty$, this is formalised by the Fuchs theorem that tells us precisely how  $\b-\b_{\sst{[0]}} \rightarrow 0$ when $\tau \rightarrow + \infty$, see the text. Note that here we consider a non--chaotic system and that the `fundamental chamber' determined by the walls in \emph{not} contained within the light cone.} }
\hspace{4.5cm} 
\begin{picture}(20,240)
\includegraphics{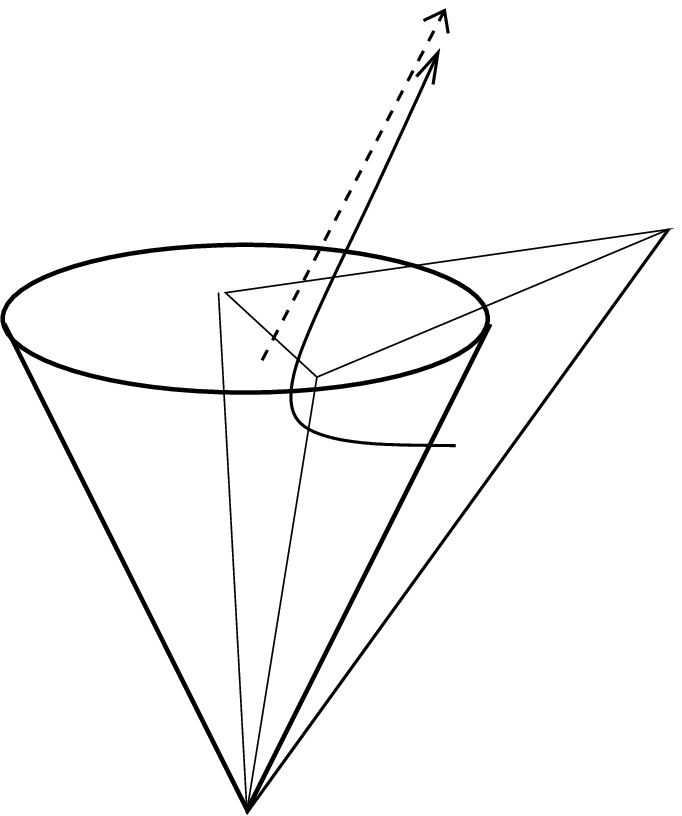}
\end{picture}
\end{figure}

\emph{\textbf{Summary}:  An asymptotic solution $\{\b_{\sst{[0]}}(\tau,x),\pi_{\sst{[0]}}(\tau,x),Q_{\sst{[0]}}(\tau,x),P_{\sst{[0]}}(\tau,x)\}$ 
(\ref{asym_sol}) obeying the asymptotic constraints (\ref{Casym}), the asymptotic evolution system (\ref{asym_eq}) and conditions (\ref{condinit}), parametrizes a solution $\{\b(\tau,x),\pi(\tau,x),Q(\tau,x),P(\tau,x)\}$ (\ref{fullsol}) of \, the full \, constrained Einstein--matter equations (this is pictured in Figure \ref{kasnerlikecase}). Moreover, the asymptotic closeness of the two solutions satisfies inequalities of the type,}
\beq
\bar \b (\tau,x)= \b - \b_{\sst{[0]}} &=&  O(e^{-\m^{(-)}\tau}) \, ,  \nn \\
\bar Q (\tau,x)= Q - Q_{\sst{[0]}} &=&O(e^{-\m^{(-)}\tau})  \, ,  \nn \\
\bar \pi (\tau,x)= \pi - \pi_{\sst{[0]}} &=& O(e^{-\m^{(-)}\tau}) \, , \nn \\
\bar P (\tau,x)= P - P_{\sst{[0]}} &=& O(e^{-\m^{(-)}\tau})   \, ;\nn \\ 
\eeq
\emph{where $\m$ is any number strictly smaller $\forall x \in U$ and $\forall A$ than the quantities $2w_A(p_\circ(x))> 2 \epsilon$, when $\tau \rightarrow + \infty$  }.\newline

$ \, $ 

\textbf{\emph{On the generality of the construction}}\newline

An important fact to notice about the above construction is that the solution of the asymptotic equations (\ref{asym_sol}) subject to the constraints (\ref{Casym}), (\ref{Caasym}) and asymptotic Gauss constraints possesses as many arbitrary functions as one expects to be present in the `general solution' of the constrained Einstein--matter equations. The inequalities (\ref{condinit}) impose restrictions on these arbitrary functions but do not change their number. Let us repeat that this construction applies only in cases where the fundamental chamber in $\b$--space defined by the inequalities $w_A(\b)\geq 0$ extends beyond the light cone $G_{ab} \b^a \b^b \leq 0$, $\sum \b >0$ as illustrated in Figure \ref{kasnerlikecase}.\newline

\textbf{\emph{Comparison with the asymptotically velocity dominated system}} \newline

The appendix \ref{appB} compares the \emph{velocity dominated system }of \cite{Andersson:2000cv,Damour:2002tc} and our approach. The essential differences between the two approaches are the following: (i) they do not use the same asymptotic system, (ii) the use of Iwasawa variables allows for a more transparent treatment of the `source terms' in the Fuchsian equations (indeed, in Iwasawa variables, it suffices to read off the exponential terms in equations (\ref{fuchsnonchaotic})), (iii) the use of Iwasawa variables avoids the technical problems linked to measuring the `difference' between the exact and asymptotic metrics by writing $(g_{[0]}^{-1}g)^a{}_b = \d^a{}_b + t^{\a^a{}_b}\g^a{}_b$ with some carefully chosen $\a$'s and an \emph{asymmetric} matrix $\g$.\newline

We should, however, remark that the two methods differ in the extension of the open regions $U$ where the Fuchsian method can be used to construct the metric. In the method of  \cite{Andersson:2000cv,Damour:2002tc} one can cover the full analytic manifold by using many small neighborhoods in which the frame approximately (but analytically) diagonalises the second fundamental form $K$. In the Iwasawa approach one can work in large open domains, but there is a problem connected with the presence of co--dimension 2 submanifolds where two eigenvalues of $K$ coincide. This problem is briefly discussed in appendix \ref{appB}. More work is needed to extend the Iwasawa--variable approach so as to be able to cover the full analytic manifold.


\section{Iwasawa--variables treatment of chaotic systems \label{chaos}}

We now turn to our main purpose, which is to give a precise formulation of the asymptotic BKL behavior in the chaotic case. In this perspective, we follow the same strategy as for the non--chaotic case: 
\begin{itemize}
\item[(i)] We first define an \emph{asymptotic evolution system, made of ordinary differential equations }(ODEs) that will describe the generic asymptotic unconstrained dynamics of the Iwasawa variables near a spacelike singularity. Of course, the solutions $\{\b_{\sst{[0]}},\pi_{\sst{[0]}},Q_{\sst{[0]}},P_{\sst{[0]}} \}$  of the asymptotic system are much more involved than a Kasner--like behavior and cannot be given in a closed form.
\item[(ii)] We then define \emph{asymptotic constraints }whose vanishing is preserved by the above defined asymptotic evolution system. 
\item[(iii)] Next, we construct a `\emph{generalized Fuchsian system}' that describes the behavior of the differences $\bar \b = \b -\b_{\sst{[0]}},\, \bar \pi = \pi -\pi_{\sst{[0]}},\, \bar Q = Q -Q_{\sst{[0]}}$ and $ \bar P = P - P_{\sst{[0]}} $ between an exact solution $\{\b,\pi,Q,P\}$ of the considered Einstein--matter system, and a solution $\{\b_{\sst{[0]}},\pi_{\sst{[0]}},Q_{\sst{[0]}},P_{\sst{[0]}} \}$ of the asymptotic evolution system. We then argue that, given a solution $\{\b_{\sst{[0]}},\pi_{\sst{[0]}},Q_{\sst{[0]}},P_{\sst{[0]}} \}$  of the asymptotic evolution system, there exists a \emph{unique} solution $\{\bar \b,\bar \pi, \bar Q, \bar P\}$ of the differenced system which goes to zero as $\tau \rightarrow + \infty$. 
\item[(iv)] We formally show that, if the asymptotic constraints are satisfied, the full constraints satisfy a generalized Fuchs system. We then argue that the full constraints will be satisfied as a consequence of the vanishing of the asymptotic ones.  
\end{itemize}
Finally, our methodology suggests that one can indeed parametrize a solution of the full constrained Einstein--matter system by a solution of the system of ODEs defined in (i). \newline

\emph{Note: }we use the same notation for the solution of the asymptotic system of equations, asymptotic solutions, asymptotic Hamiltonian etc in the chaotic and non--chaotic cases. 


\subsection{Definition of the asymptotic evolution equations}

The billiard picture provides a guide for choosing a suitable asymptotic evolution system since it gives us an intuitive description of the asymptotic dynamics. In the billiard approximation, the dynamics of the  `diagonal variables' $\b$'s is described as a free Lorentzian motion interrupted by reflections upon infinite--potential walls and the `off--diagonal variables' $Q$'s are frozen. Here, we shall go beyond this simplified `sharp wall' billiard picture and work with exponential (`Toda') potential walls. The asymptotic system should completely determine the asymptotic dynamics, given some suitable initial data. It is therefore crucial to use a system of \emph{ordinary} differential equations (rather than partial differential equations) to characterize the asymptotic dynamics, so that we are able to use theorems about existence and uniqueness of solutions. \newline

Motivated by these reasons, we \emph{define }(for any chaotic Einstein--matter system) an asymptotic evolution system in the following way: for the `diagonal variables', we keep in the Einstein--matter equations (\ref{general_eq}) only the dominant exponential walls while for the `off--diagonal variables', we neglect all the walls in the equations of motion.
These prescriptions define an `asymptotic evolution system' which reads, in sketchy form:
\beq 
\partial_\tau \beta_{\sst{[0]}} &=& {1 \over 2}\pi_{\sst{[0]}} \nn \\ 
\partial_\tau \pi_{\sst{[0]}} &=& \sum_\ca 2 c_{\sst{\ca}}(Q,P,\pa_x Q) w_{\sst{\ca}} e^{-2 w_{\sst{\ca}}(\b_{\sst{[0]}})} 
\nn \\ 
\partial_\tau Q_{\sst{[0]}} &=& 0 
\nn \\
\partial_\tau P_{\sst{[0]}} &=& 0 \, .
\label{eq_asympt_chaos}
\eeq
Here $\ca$ labels the dominant walls only. As exhibited in equation (\ref{potentiel}) the coefficients $c_\ca$ of the dominant Toda walls depend only on the $Q$'s, $P$'s and the first spatial derivative of the $Q$'s. It is crucial that they do not involve the \emph{spatial derivatives of the} $\b$'s. The solutions of  the last two equations in the system (\ref{eq_asympt_chaos}) are simply $Q_{\sst{[0]}} = Q_{\sst{(0)}}(x)$, $P_{\sst{[0]}} = P_{\sst{(0)}}(x)$. Considering $Q_{\sst{(0)}}(x)$ and $P_{\sst{(0)}}(x)$ as given data, and replacing them in the other equations of the system (\ref{eq_asympt_chaos}), we see that the diagonal variables $\{\b_{\sst{[0]}},\pi_{\sst{[0]}}\}$, at each given spatial point $x$, satisfy a system of ODEs. It is then easily checked that the latter system of ODEs follows from the Hamiltonian $\ch_\circ$:
\beq 
\ch_{\circ \, Q_{\sst{(0)}},P_{\sst{(0)]}}}[\beta_{\sst{[0]}}; \pi_{\sst{[0]}}] = {1\over 4} G^{ab} \pi_{\sst{[0]}\, a} \pi_{\sst{[0]} \, b} 
+ \sum_{\sst{\ca}} c_{\sst{\ca}}(P_{\sst{(0)}},Q_{\sst{(0)}},\pa_a Q_{\sst{(0)}}) e^{-2w_{\sst{\ca}}(\b_{\sst{[0]}})}\label{asym_ham_c} 
 \, .
\eeq  
Note that, from \cite{Damour:2002et}, the qualitative behavior (as $\tau \rightarrow + \infty$) of the solution for such systems of equations is as follows:  
the $\pi_{\sst{[0]}}$'s go to zero\footnote{Actually, this property is guaranteed only if one imposes the constraint $\ch_{\circ \, Q_{\sst{(0)}} P_{\sst{(0)}} } = 0$.} and the $\b_{\sst{[0]}}$'s behave approximately as in the sharp billiard picture (free motions `$p \tau + const$' interrupted by collisions against the `walls' $e^{-2w_{\sst{\ca}}(\beta)}$).


\subsection{Definition of the asymptotic constraints}

It is natural to define the asymptotic Hamiltonian constraint to be
\beq
\ch_{\sst{[0]}} := {1\over 4} G^{ab} \pi_{\sst{[0]}\, a} \pi_{\sst{[0]}\, b}
+ \sum_{\sst{\ca}} c_{\sst{\ca}}(P_{\sst{[0]}},Q_{\sst{[0]}},\pa_a Q_{\sst{[0]}}) e^{-2w_{\sst{\ca}}(\b_{\sst{[0]}})} \, . \label{constcchaos} 
\eeq 
Like in the non--chaotic case the asymptotic Hamiltonian constraint (\ref{constcchaos}) coincides with the asymptotic evolution Hamiltonian (\ref{asym_ham_c}), and is therefore preserved by the asymptotic time evolution. 
Let us now define the asymptotic momentum constraints as the formal limit of the `full' momentum constraints when $\tau \rightarrow + \infty$. We start from the general expression (\ref{momconstgeneral}) for the momentum constraints. In view of what was recalled from \cite{Damour:2002et}, $\pi \rightarrow 0$ as $\tau \rightarrow + \infty$ (after imposing (\ref{constcchaos})), so that we can discard the terms linear in $\pi$ in the previously considered non--chaotic asymptotic momentum constraints (\ref{Caasym}) (which neglected exponential walls that we still formally neglect). We therefore define the asymptotic momentum constraints as follows,
\beq
-{1 \over 2}\ch_{a \sst{[0]}} &:=& 
\big[\pa_b 
\pi_{\sst{\mathrm{iwa}}}{}^b{}_{a \, [-]}
+ C_{\sst{\mathrm{iwa}}}{}^c{}_{cb} 
\pi_{\sst{\mathrm{iwa}}}{}^b{}_{a \,[-]}
+ C_{\sst{\mathrm{iwa}}}{}^d{}_{ac} 
\pi_{\sst{\mathrm{iwa}}}{}^c{}_{d \, [-]} \nn \\
& & + {1 \over p!} \ce^{a_1...a_p} \cf^{\sst{(p)}}_{a a_1...a_p} \big]^{[0]}\, , \hspace{1cm} \mathrm{no \, sum \, over \, a, \, sum \, over \, d}\label{constcachaos} 
\eeq
where the $\cf^{\sst{(p)}}_{a a_1...a_p}$'s are the $F^{\sst{(p)}}_{\sst{(i)(i_1)...(i_p)}}$'s expressed in the Iwasawa basis, where $\pi_{\sst{[0]}}{}^b{}_{a\, \sst{[-]}}$ is defined as the r.h.s. of the second equation (\ref{pivsp}) and where the overall bracket $[\, ]^{\sst{[0]}}$ means that one must do the replacements $Q \rightarrow Q_{\sst{[0]}}, \, P \rightarrow P_{\sst{[0]}}$. Note that the momentum constraint (\ref{constcachaos}) contains only the time--independent quantities $\cn_{\sst{[0]}} , \pa_x \cn_{\sst{[0]}}, Q_{\sst{[0]}}, P_{\sst{[0]}}$ and therefore is trivially preserved by the time evolution. \newline

Finally, the asymptotic Gauss constraint for each $p$--form is defined to be the Gauss constraint with the asymptotic variables $Q_{\sst{[0]}}, \, P_{\sst{[0]}} $ instead of $Q, \, P$, \ie 
 \beq 
\varphi^{a_1... a_{p-1}}_{\sst{(p) \, [0]}} &: =& \pa_{a_p} \pi_\so{}^{a_1...a_p}
-{1\over 2} C_{\sst{\mathrm{iwa}} \so}{}^{a_1}{}_{b a_p}  \pi_\so{}^{ba_2...a_p} -... - 
{1\over 2} C_{\sst{\mathrm{iwa}} \so}{}^{a_{p-1}}{}_{b a_p}  \pi_\so{}^{a_1...ba_p} \nn \\
& &+  C_{\sst{\mathrm{iwa}} \so}{}^{a_p}{}_{a_p b} \pi_\so{}^{a_1...a_{p-1}b} \, .
\label{agcc}
\eeq
These constraints are preserved by the asymptotic time evolution since the $\pi_\so{}^{a_1...a_p}$ (which is one of the $P_{\sst{[0]}} $'s) and $C_{\sst{\mathrm{iwa}} \so}{}^{a}{}_{b c}$ (which depends on the $\cn_\so \in Q_{\sst{[0]}}$'s via 
(\ref{iwastructurefunctions})),  are constants according to the asymptotic evolution equations (\ref{eq_asympt_chaos}). 


\subsection{Construction of a `generalized Fuchsian' system for the `differenced variables' \label{genefuchs}}

We now rewrite equations (\ref{general_eq}) in terms of the differences $\{\bar{\b},\, \bar \pi, \, \bar Q, \, \bar P\}$,  
\beq
\beta^a  &=& \beta_{\sst{[0]}}^a + \bar \beta^a \, , \nn \\
\pi_a  &=& \pi_{\sst{[0]}\, a} +\bar \pi_a \, , \nn \\
Q  &=& Q_{\sst{[0]}} + \bar Q\, , \nn \\
P &=& P_{\sst{[0]}} + \bar P\, .
\label{fullsolchaos}\eeq 
This gives (when suppressing indices on $\bar \b$ and $\bar \pi$),
\beq 
\partial_\tau  \bar \beta - {1\over 2}\bar \pi &=& 0 \nn \\
\partial_\tau  \bar \pi &=& 
2  \sum_{\sst{\ca}}  w_{\sst{\ca}}  e^{-2 w_{\sst{\ca}} (\beta_{\sst{[0]}})} 
(c_{\sst{\ca}}e^{- 2w_{\sst{\ca}}(\bar \beta)}-c_{\sst{\ca}}(
Q_{\sst{[0]}},P_{\sst{[0]}},\partial_x Q_{\sst{[0]}})) \nn \\ 
&+&
2  \sum_{\sst{\ca^\prime}} c_{\sst{\ca^\prime}} w_{\sst{\ca^\prime}}  e^{-2w_{\sst{\ca^\prime}}(\beta_{\sst{[0]}})} e^{- 2w_{\sst{\ca^\prime}}(\bar \b)} \nn \\ 
&+&
  \sum_{\sst{\ca^\prime}} \partial_x ( { \partial c_{\sst{\ca^\prime}} \over \partial \partial_x \b} e^{-2_{\sst{\ca^\prime}} (\beta_{\sst{[0]}} ) }e^{- 2w_{\sst{\ca^\prime}}(\bar \beta)} ) \nn \\
&-&
 \sum_{\sst{\ca^\prime}} \partial^2_x ( { \partial c_{\sst{\ca^\prime}}\over \partial \partial^2_x \b} e^{-2w_{\sst{\ca^\prime}} (\beta_{\sst{[0]}} ) }e^{-  2w_{\sst{\ca^\prime}}(\bar \beta)} ))
\nn \\
\partial_\tau  \bar Q &=& \sum_\ca {\partial c_\ca \over \partial P} e^{-2w_{\ca} (\beta_{\sst{[0]}} ) }e^{- 2w_{\ca}(\bar \beta)}\nn\\
&+&\sum_{\ca^\prime}{\partial c_{\ca^\prime} \over \partial P} e^{-2w_{\ca^\prime} (\beta_{\sst{[0]}} ) }e^{- 2w_{\ca^\prime}(\bar \beta)}\nn\\
\partial_\tau \bar P &=& \sum_{\ca} \left(-{\partial c_{\ca}\over \partial Q} e^{-2w_{\ca}} + 
\partial_x({\partial c_{\ca} \over \partial \partial_x Q} e^{-2w_{\ca} (\beta_{\sst{[0]}} ) }e^{- 2w_{\ca}(\bar \beta)})\right) \nn \\
&+&\sum_{\ca^\prime} \left(-{\partial c_{\ca^\prime}\over \partial Q} e^{-2w_{\ca^\prime}} + 
\partial_x({\partial c_{\ca^\prime} \over \partial \partial_x Q} e^{-2w_{\ca^\prime} (\beta_{\sst{[0]}} ) }e^{- 2w_{\ca^\prime}(\bar \beta)})\right) \nn \\
&-&
\sum_{\ca^\prime}\partial^2_x({\partial c_{\ca^\prime} \over \partial \partial^2_x Q} e^{-2w_{\ca^\prime} (\beta_{\sst{[0]}} ) }e^{- 2w_{\ca^\prime}(\bar \beta)})\, 
\, ,\label{fuchschaotic} \eeq
where we recall that $\ca$ labels the dominant exponential walls (with coefficient $c_\ca$ which depend only on $\{P,Q,\pa_a Q\}$) while ${\ca^\prime}$ labels the subdominant exponential walls (with  coefficient $c_{\ca^\prime}$ which depend on  $\{P,Q,\pa_a Q,\pa^2_a Q, \pa_a  \b, \pa^2_a \b \}$). In addition, in all the coefficients $c_\ca$ and $c_{\ca^\prime}$ on the r.h.s. of the system (\ref{fuchschaotic}) one must do the replacements (\ref{fullsolchaos}), so that for us 
$c_\ca=c_\ca(Q_{\sst{[0]}}+\bar Q,P_{\sst{[0]}}+ \bar P,\partial (Q_{\sst{[0]}}+\bar Q))$
and
$c_{\ca^\prime}=c_{\ca^\prime}(Q_{\sst{[0]}}+\bar Q,P_{\sst{[0]}}+ \bar P,\partial_x (\b_{\sst{[0]}} + \bar \beta ), \partial^2_x  (\b_{\sst{[0]}}+ \bar \beta ) ,\partial (Q_{\sst{[0]}}+\bar Q), \partial^2 (Q_{\sst{[0]}}+\bar Q )) $. \newline

This system of equations is not a Fuchsian system as defined above. However, it is similar to such a system. Indeed, it contains a space and time independent matrix $\ca$ which is again given by (\ref{matrixA}). However, the crucial difference between (\ref{fuchschaotic}) and a Fuchsian system concerns the source term on the r.h.s.. Instead of containing (modulo a bounded term) a space--independent factor which is exponentially decreasing with $\tau$, $e^{-\m \tau}$, it contains exponential wall terms $e^{-2 w_{\ca}(\b_{ \sst{[0]} } )}$ and $e^{-2w_{\ca^\prime}(\b_{\sst{[0]}})}$ where $\b_{\sst{[0]}}(\tau,x)$ is a solution of the asymptotic evolution system (\ref{eq_asympt_chaos}).
\newline

Let us qualitatively analyze the behavior of this source term as $\tau \rightarrow+ \infty$. For that, let us start by recalling the sketchy time dependence of the solution $\b_{\sst{[0]}}$ that would be given by the billiard picture: namely a succession of Kasner epochs. During each Kasner epoch the source term is exponential decreasing (in the time coordinate $\tau$), indeed, during the `Kasner free motion' $\b_{\sst{[0]}} \approx p_\circ \tau + \b_\circ$ so that each exponential wall factor term $e^{-2w_A(\b_{\sst{[0]}})}$ ($A = \ca,\ca^\prime$) behaves has 
$e^{-\m (x) \tau}$ with $\mu(x) \approx 2 w_A(p_\circ(x))$. However, this exponential decrease is interrupted around the instants of collision on the dominant walls, during which $w_\ca(\b_{\sst{[0]}})$ in fact vanishes so that $e^{-2 w_A(\b_{\sst{[0]}})}$ would seem to become unity. 
By contrast the subdominant terms $e^{-2w_{\ca^\prime}(\b_{\sst{[0]}})}$ are always exponentially decreasing because the $\b$--particle generically never hits them, being deflected by a collision on dominant walls before reaching them. [We are here neglecting the measure zero set of trajectories which exactly hit a `corner' of the billiard, where a subdominant wall intersects dominant ones.]  \newline

The sharp billiard picture just recalled is only an approximation to the asymptotic dynamics (\ref{eq_asympt_chaos}). When taking into account the existence of exponential walls in the equations (\ref{eq_asympt_chaos}) one can describe more precisely the behavior of $(\b_{\sst{[0]}}, \pi_{\sst{[0]}})$ and thereby of the crucial `source terms' $\propto e^{-2w_A(\b_{\sst{[0]}})}$ appearing on the r.h.s. of (\ref{fuchschaotic}). Indeed, following the method used in \cite{Damour:2002et} one can conveniently analyze the dynamics of $(\b_{\sst{[0]}}, \pi_{\sst{[0]}})$ following from the asymptotic Hamiltonian $\ch_{\sst{[0]}}$ (\ref{asym_ham_c}), and submitted to the zero--energy constraint (\ref{constcchaos}). When decomposing $\b^a$ as $\b^a = \rho \g^a$ with $\rho^2 =-G_{ab} \b^a \b^b$ and $G_{ab} \g^a \g^b = -1$, which correspondingly implies $\pi_a =  \rho^{-1} \, \pi^\g_a - \pi_\rho \g_a$ where $\pi^\g_a$ (submitted to the constraint $\g^a \pi_a^\g =0 $ ) is conjugate to the `position' $\g^a$ on the unit hyperboloid ($G_{ab}\g^a \g^b=-1$), and where $\pi_\rho$ is the conjugate to the variable $\rho$, one finds that the Hamiltonian reads
\beq
\ch_{\sst{[0]}} = {1 \over 4}Ê(- \pi_\rho^2 + { \pi_\g^2 \over \rho^2} )+ \cv_{\sst{[0]}} \, , \label{hamgamma}
\eeq
where
\beq
\cv_{\sst{[0]}} (\rho,\g) := \sum c_\ca e^{-2 \rho w_A(\g)} \, . 
\eeq
The zero--energy constraint can then be written as 
\beq
 {1 \over 4}Ê( - \pi_\lambda^2 + \pi_\g^2) + \rho^2 \cv_{\sst{[0]}} =0\, , \label{constgamma}
\eeq
where $\pi_\lambda := \rho \pi_\rho$ is now conjugate to $\lambda := \ln{\rho}$. From this constraint one infers (see \cite{Damour:2002et}) that, as $\tau \rightarrow + \infty$ and therefore $\rho \rightarrow + \infty$, $\pi_\lambda$ tends to a finite limit say $p_\lambda$, and therefore $\arrowvert \pi_\g \arrowvert$ oscillates between $p_\lambda$ (far from the walls) and 0 (during a `collision'). From this result one also infers that the maximum value of $\ch_{\sst{[0]}} $, reached during a `collision' (\ie when $\arrowvert \pi_\g \arrowvert = 0$), is such that 
$\rho^2 \cv_{\sst{[0]}} = {1\over 4} \pi_\lambda^2 \rightarrow {1 \over 4}Êp_\lambda^2$ as $\tau \rightarrow + \infty$. Finally, one concludes that as $\tau \rightarrow + \infty$, and therefore $\rho \rightarrow + \infty$ (roughly proportionally to $\tau$) even the maximum values of the dominant exponential potential (reached during the collision) decay like $\cv_{\sst{[0]}} \propto \rho^{-2}$. [One also conclude from $\pi_a =  \rho^{-1} \, (\pi_a^\g - \pi_\lambda \g_a)$ that the components of the $\b$--conjugate momenta $\pi_a $ decay proportionally to $\rho^{-1}$.] 
\newline

\begin{figure}[h]
\begin{flushleft}
 \begin{picture}(0,0)%
 \includegraphics{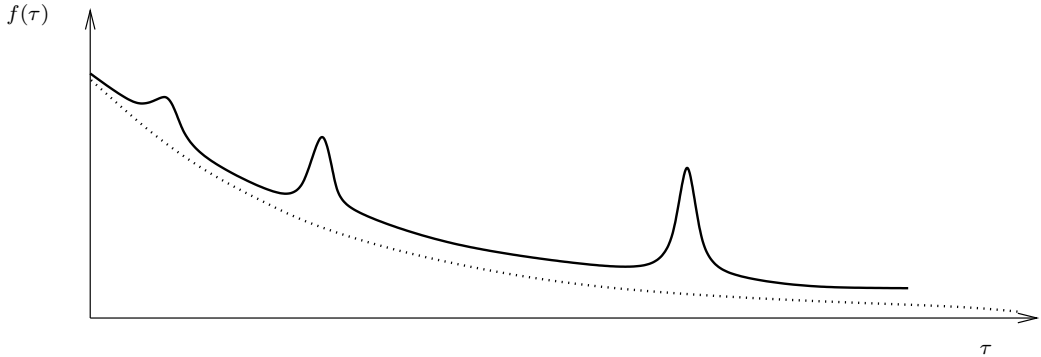}%
 \end{picture}%
\setlength{\unitlength}{1973sp}%
\begingroup\makeatletter\ifx\SetFigFont\undefined%
\gdef\SetFigFont#1#2#3#4#5{%
  \reset@font\fontsize{#1}{#2pt}%
  \fontfamily{#3}\fontseries{#4}\fontshape{#5}%
  \selectfont}%
\fi\endgroup%
\begin{picture}(12997,4455)(151,-3874)
\put(12376,-3811){\makebox(0,0)[lb]{\smash{\SetFigFont{8}{9.6}{\familydefault}{\mddefault}{\updefault}{\color[rgb]{0,0,0}$\tau$}%
}}}
\put(151,389){\makebox(0,0)[lb]{\smash{\SetFigFont{8}{9.6}{\familydefault}{\mddefault}{\updefault}{\color[rgb]{0,0,0}$f(\tau)$}%
}}}
\end{picture}
\end{flushleft}
 \caption{\small{Schematic drawing of the source term of the system of equations (\ref{fuchschaotic}). }}\label{source}
\end{figure}

Summarising, we conclude that each dominant potential term $e^{-2 w_\ca(\b_\so )}$ entering the r.h.s. of (\ref{fuchschaotic}) has the qualitative behavior depicted in Figure \ref{source}, namely an overall exponential decay, interrupted by `peaks' (of decreasing magnitude $\propto \rho^{-2}$) corresponding to collisions. In addition, (see appendix A of \cite{Damour:2002et}) the $\tau$--time spacing between successive peaks increases (roughly like $\ln{\rho}  \sim \ln{\tau}$) as $\tau \rightarrow + \infty$. As for the behavior of the subdominant exponential potential terms $e^{-2 w_{\ca^\prime}(\b_\so)}$ entering the r.h.s. of (\ref{fuchschaotic}) it is expected to be somewhat similar to the one depicted in Figure \ref{source}, except for the facts that the overall exponential decay should be faster, and that the peaks should be much rarer (corresponding to a collision happening nearly in a `corner'). [We expect that the faster decay and the rarer occurrence of peaks also compensates the fact that the presence of $(\pa_x \b)^2$ and $\pa_x^2 \b$ in the coefficient $c_{\ca^\prime}$ generates a growing behavior $\propto \tau^2$ of the $c_{\ca^\prime}$.]  

The system (\ref{fuchschaotic}) for $\{\bar \b, \bar \pi, \bar Q,\bar P\}$ (in which the r.h.s. depends on a solution $\{ \b_\so,  \pi_\so,$ $  Q_\so, P_\so \}$ of the asymptotic system (\ref{eq_asympt_chaos})), can be viewed as a \emph{generalized Fuchs system}. Note that the structure of this `generalized Fuchs system' is of the form 
\beq 
\pa_\tau u - \ca u = \sum_A e^{-2 w_A(\b_\so)} f_A(\b_\so, \pa_x \b_\so,\pa_x^2 \b_\so, x,u,\pa_x u, \pa_x^2 u) \, , \label{genfuchs}
\eeq
where $u$ is a vector--valued unknown function $u(\tau,x) = (u_1(\tau,x), ...,u_k(\tau,x))$, the linear forms $w_A(\b)$ are the same ones that enter in the system (\ref{fuchschaotic}) and where the source terms $f_A$ can be read off the system (\ref{fuchschaotic}). In view of the arguments given in \cite{Damour:2002et} (in particular, we recall that in the appendix A of this reference, it has been argued that the `peaks' in the source terms pictured in Figure \ref{source} are such that their integrated effect allows $u$ to have a limit as $\tau \rightarrow + \infty$) and partially recalled above, we expect that (under the conditions specified below) there exists a \emph{unique} solution $\{ \bar \b, \bar \pi, \bar Q, \bar P\}$ of (\ref{fuchschaotic}) tending to zero as $\tau \rightarrow + \infty$ (and more generally a \emph{unique} solution $u$ of the system (\ref{genfuchs}), given suitable conditions on $f$, which tends to zero as $\tau \rightarrow + \infty$).  
\newline

The conditions necessary for this result to hold for (\ref{fuchschaotic}) are expected to be the following: 
\begin{itemize}
\item the asymptotic initial data $Q_\circ (x), P_\circ(x)$ must be such that the coefficients $c_\ca(Q_\circ (x),$ $P_\circ(x), \pa_x Q_\circ (x))$ of the dominant potential walls remain strictly positive over the considered domain $U$, 
\item the asymptotic initial data $\b_\circ(x)=\b(\tau_1,x),\, \pi_\circ(x)=\pi(\tau_1,x)$ at some finite time $\tau_1$ must satisfy the zero--energy constraint $\ch_\so(\b_\circ,\pi_\circ,Q_\circ,P_\circ)=0$.
\end{itemize}


\subsection{Constraints}

To complete the story, we need to check that the exact constraints are satisfied (along the exact equations of motion) once the asymptotic ones are fulfilled (along the asymptotic equations of motion). The reasoning is the same as in the non--chaotic case: 
\begin{itemize}
\item[1.] First we treat the Gauss constraints. We impose that the asymptotic Gauss constraints (\ref{agcc}) hold. It is again obvious that these asymptotic constraints are preserved by the asymptotic equations of motion (\ref{eq_asympt_chaos}).  On the other hand, the exact Gauss constraints (\ref{Gauss}) are preserved by the exact evolution equations. Moreover, they differ from the asymptotic ones by exponential `walls', and consequently they vanish when $\tau \rightarrow + \infty$. As in the non--chaotic case, we can conclude that they are equal to zero (because they are constant and they tend to zero). 
\item[2.] We then turn to the Hamiltonian and momentum constraints.  Let us require that the initial data $Q_\circ, \, P_\circ$ of the asymptotic evolution system satisfy the asymptotic momentum constraints $\ch_{\so a} = 0$, (\ref{constcachaos}) in addition to the asymptotic Hamiltonian constraint $\ch_{\sst{[0}]}=0$. The constraints obey the evolution system (\ref{evoexconst}),\footnote{See appendix \ref{hamform} and the discussion in the section about the constraints in the non--chaotic case} \ie 
\beq
 \partial_\tau \ch &=& Ê\sum_a e^{-2 \mu_a(\b)} (\pa_a \ch_a + 2 \b^a_{\, , a} \ch_a -2(\sum_c \b^c)_{,a} \ch_a)\, , \nn \\
 \partial_\tau \ch_a - \nabla_a\ch &=& 0
\, . \label{evoexconstbis}
\eeq
As in the non--chaotic case, the above system is not Fuchsian due to (i) the term $\nabla_a \ch$ [however, as we have argued previously, this term should not be a problem], (ii) the source term (\ie the r.h.s. of the first equation in the system (\ref{evoexconstbis})) which is not an allowed Fuchsian source term (this fact contrasts with the non--chaotic case).  However we have already dealt with this kind of source term in the system (\ref{fuchschaotic}). In the present case, the source term is even `better' than the one of (\ref{fuchschaotic}) because it contains only `subdominant walls' which decay faster than the dominant ones and which exhibit rearer `peaks' (because the `peaks' occur when the `ball' hits a corner) [however these `peaks' contain a factor $\propto \tau^2$] (see discussion in section \ref{genefuchs}). Accordingly, the system (\ref{evoexconstbis}) is a `generalized Fuchsian' system in the sense given in section \ref{genefuchs} and we consider it likely that it possesses a \emph{unique }Êsolution that vanishes when $\tau \rightarrow + \infty$. On top of that, the system (\ref{evoexconstbis}) is homogeneous and this implies that the unique solution in question is zero.  \newline

We can then conclude that the exact Hamiltonian and momentum constraints are satisfied since (i) they differ from the asymptotic ones, which are imposed to hold, by exponential `walls' (this implies that they decay as $\tau \rightarrow + \infty$), (ii) they obey a `generalized Fuchs system', as just argued, and therefore there exists a \emph{unique }Êsolution that vanishes when $\tau \rightarrow + \infty$. 
\end{itemize}

\emph{\textbf{Summary}}:
Let us summarise our conjectural results concerning the asymptotic dynamics of the fields in the vicinity of a spacelike singularity for a chaotic Einstein--matter systems in the following statement: \newline

Let 
\begin{itemize}
\item  $( Q_{\sst{[0]}}(x),P_{\sst{[0]}}(x))$ be functions of the spatial coordinates such that the coefficients 
$c_A(Q_{\sst{[0]}}(x),P_{\sst{[0]}}(x),\pa_a Q_{\sst{[0]}}(x))$ nowhere vanish, and that the `fundamental chamber' defined by the inequalities $w_\ca(\b) \geq 0$ is contained within the future lightcone $G_{ab} \b^a \b^b=0$, $\sum_a \b^a \geq 0$, 
\item $(\b_{\sst{[0]}},\pi_{\sst{[0]}})$ be a solution a the asymptotic system of equations (\ref{eq_asympt_chaos}) with initial conditions $(\b_{\sst{[0]}}(x),\pi_{\sst{[0]}}(x))$ given at some finite $\tau =\tau_1$ which satisfy the asymptotic Hamiltonian constraint (\ref{constcchaos}),  
and
\item impose that the asymptotic momentum constraints (\ref{constcachaos}) are satisfied at $\tau = \tau_1$ as well as the asymptotic Gauss constraints (\ref{agcc}).
\end{itemize}
Then there exists a unique solution $(\b,\pi,Q,P)$ of the Iwasawa--variable form of the full constrained Einstein--matter equations such that the differences $\bar \b = \b - \b_{\sst{[0]}} \,\bar \pi = \pi - \pi_{\sst{[0]}} \,\bar Q = Q - Q_{\sst{[0]}}, \bar P = P - P_{\sst{[0]}} \}$ tend to zero as $\tau \rightarrow + \infty$. 
\begin{figure}[h]
 \caption{ \label{nonkasnerlikecase} \small{ \emph{Chaotic behavior } This picture is a schematic drawing of the asymptotic dynamics of the `diagonal variables' at a given spatial point $x$, this dynamics is represented in the $\b$--space. The dashed curve represents the zeroth order solution $\b_{\sst{[0]}}$. The exact solution is sketched as a continuous curve. The idea is that the approximate solution $\b_{\sst{[0]}}$ becomes better and better as $\tau \rightarrow + \infty$, this is formalised via a `generalized Fuchs theorem', see the text. Note that here we consider a chaotic system and that the `fundamental chamber' determined by the walls in contained within the light cone.} }
\hspace{4.5cm} 
\begin{picture}(20,240)
\includegraphics{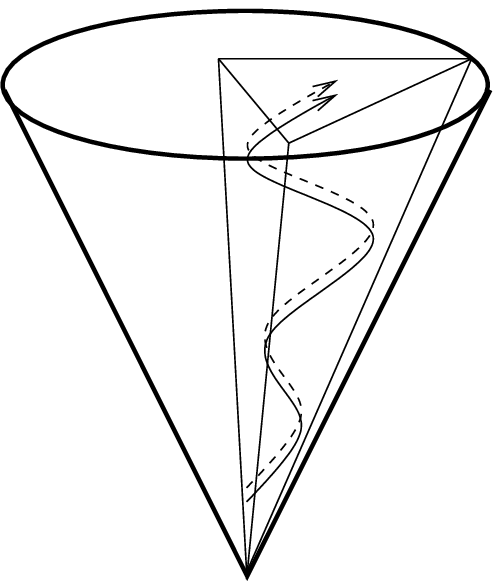}%
\end{picture}%
\end{figure}


\section{Pure gravity in dimensions $4 \leq D \leq 10$}

To give a more concrete example of the general formulation of the BKL behavior of Einstein--matter systems discussed here, let us consider the specific example of pure gravity. We consider spacetime dimensions $D$ such that $4 \leq D \leq 10$, so that the corresponding behavior is generically chaotic. For this case, our precise formulation of the BKL conjecture is the following (we denote by $d := D - 1$ the space dimension $3 \leq d \leq 9$):
\begin{itemize}
\item[(i)] \textbf{Initial data} Let us give ourselves the following initial data: \newline
$d(d-1)/2$ spatial functions $\cn_{\sst{(0)}}{}^a{}_i (x) $ for $a <i$, \newline
$d(d-1)/2$ spatial functions $\cp_{\sst{(0)}}{}^i{}_a (x) $ for $a <i$, \newline 
$d$ spatial functions $\b_{\sst{(0)}}{}^a (x)$ and \newline
$d$ spatial functions $\pi_{\sst{(0)}Ê\, a} (x)$.
\item[(ii)] \textbf{Asymptotic Hamiltonian} Given these data we define the following asymptotic Hamiltonian 
\begin{equation}
\label{eq9.6}
{\mathcal H}^{\rm asymp} (\beta_\so , \pi_\so) = \frac{1}{4} \, G^{ab} \, \pi_{\so \, a} \, \pi_{\so \, b} + {\mathcal V}_S^{\rm asymp} + {\mathcal V}_G^{\rm asymp} \, ,
\end{equation}
where $G^{ab} \, \pi_{\so \, a} \, \pi_{\so \, b} := \underset{a=1}{\overset{d}{\sum}} \, \pi_{\so \, a}^2 - \frac{1}{d-1} \left( \underset{a=1}{\overset{d}{\sum}} \, \pi_{\so \, a} \right)^2$, and where
\begin{equation}
\label{eq9.7}
{\mathcal V}_S^{\rm asymp} = \frac{1}{2} \, \sum_{a=1}^{d-1} \, e^{-2(\beta_\so{}^{a+1} - \beta_\so{}^a)} ({\mathcal P}_{\sst{(0)}a}^i \, {\mathcal N}_{\sst{(0)}i}^{a+1})^2 \, ,
\end{equation}
(where $i=1,\ldots ,d$ is summed over) and
\begin{equation}
\label{eq9.8}
{\mathcal V}_G^{\rm asymp} = \frac{1}{2} \, e^{-2\alpha_{1d-1d} (\beta_\so)} (C_{\sst{(0)}d-1d}^1)^2 \, .
\end{equation}
In the last equation, $\alpha_{abc} (\beta)$ (for $b \ne c$) denotes the linear form $\alpha_{abc} (\beta) = \beta^a + \underset{e \ne b,c}{\sum} \, \beta^e$ (evaluated for $a=1$, $b=d-1$ and $c=d$), and $C_{\sst{(0)}}{}^a{}_ {bc}$ (with $b \ne c$ and $C_{\sst{(0)}}{}^a{}_{ bc} = - C_{\sst{(0)}}{}^a{}_{cb}$) denote the structure functions $(d\theta_{\sst{(0)}}{}^a = - \frac{1}{2} \, C_{\sst{(0)}}{}^a{}_{bc} \wedge \theta_{\sst{(0)}}{}^b \, \theta_{\sst{(0)}}{}^c$) of the `asymptotic Iwasawa frame' $\theta_{\sst{(0)}}{}^a (x) = {\mathcal N}_{\sst{(0)} \, i}{}^a (x) \, \omega^i$. Note that all the coefficients entering the exponential potential terms (\ref{eq9.7}) and  (\ref{eq9.8}) depend only on the spatial point (through ${\mathcal P}_{\sst{(0)}} (x)$, ${\mathcal N}_{\sst{(0)}} (x)$ and $\partial_x \, {\mathcal N}_{\sst{(0)}} (x)$ which enters $C_{\sst{(0)}}$), so that the asymptotic evolution system for $\beta$ and $\pi$ constitutes, at each point of space, a well-defined system of ODE's. 
\item[(iii)] \textbf{Asymptotic evolution equations} The equations of motion deduced from the Hamiltonian (\ref{eq9.6}) are called the \emph{asymptotic evolution equations}, they are the `chaotic analog' of the AVTD evolution system considered in the non-chaotic, monotonic power-law case. They are of the form:
\begin{eqnarray}
\label{eq9.9}
\partial_{\tau} \, \beta_{\so}^a &= &\frac{1}{2} \, G^{ab} \, \pi_{\so \, b} \, , \nonumber \\
\partial_{\tau} \, \pi_{\so \, a} &= &- \frac{\partial}{\partial \, \beta_{\so}{}^a} \, \bigl[ {\mathcal V}_S^{\rm asymp} (\beta_{\so} ; {\mathcal P}_{\sst{(0)}} , {\mathcal N}_{\sst{(0)}}) \nonumber \\
&&+ \, {\mathcal V}_G^{\rm asymp} (\beta_\so ; {\mathcal P}_{\sst{(0)}} , {\mathcal N}_{\sst{(0)}} , \partial_x \, {\mathcal N}_{\sst{(0)}}) \bigl] \,  .
\end{eqnarray}

\item[(iv)] \textbf{Asymptotic constraints} We impose that the initial data satisfy the following \emph{asymptotic constraints},
\begin{eqnarray}
\label{eq9.10}
&&{\mathcal H}^{\rm asymp} (\beta_\so , \pi_\so , {\mathcal N}_{\sst{(0)}} , \partial_x \, {\mathcal N}_{\sst{(0)}} , {\mathcal P}_{\sst{(0)}}) = 0 \, , \nonumber \\
&&{\mathcal H}_a^{\rm asymp} ({\mathcal N}_{\sst{(0)}} , \partial_x \, {\mathcal N}_{\sst{(0)}} , {\mathcal P}_{\sst{(0)}}) = 0 \, ,
\end{eqnarray}
where ${\mathcal H}^{\rm asymp}$ is the (conserved) quantity defined in equation (\ref{eq9.6}), and where the definition of ${\mathcal H}_a^{\rm asymp}$ is equation (\ref{constcachaos}) above. 
\end{itemize}
Finally, this leads to the following precise formulation of the:

\medskip

\noindent {\bf BKL conjecture in Iwasawa variables.} Let, for $x \in U$, the spatial functions
${\mathcal P}_{\sst{(0)}} \, {\mathcal N}_{\sst{(0)}}$ and $C_{\sst{(0)}}$ be such that the $d$ $x$--dependent coefficients $\cp_{\sst{(0)}}{}^i{}_a \cn_{\sst{(0)}}{}^{a+1}{}_i$ and $C_{\sst{(0)}}{}^1{}_{d-1 \, d}$ (whose squares define the coefficients of the $d$ exponential potential terms (\ref{eq9.7}) and (\ref{eq9.8})) do not vanish in $U$. 
Let $(\beta_{\so} (\tau , x) , \pi_{\so} (\tau , x)$,  be the unique solution of the asymptotic evolution system (\ref{eq9.9}) with initial conditions $\b_\so(\tau_1, x) = \b_{\sst{(0)}}(x)$ and $\pi_\so(\tau_1, x) = \pi_{\sst{(0)}}(x)$ at some finite time $\tau = \tau_1$ and  satisfying the asymptotic constraints (\ref{eq9.10}). Then there exists a unique solution $(\beta (\tau , x), \pi (\tau , x) , {\mathcal N} (\tau , x) ,$ $ {\mathcal P} (\tau , x))$ of the vacuum Einstein equations (including the constraints) such that the differences $\bar\beta (\tau , x) \equiv \beta (\tau , x) - \beta_{\so} (\tau , x)$, $\bar\pi (\tau , x) \equiv \pi (\tau , x) - \pi_{\so} (\tau , x)$, $\bar{\mathcal N} (\tau , x) \equiv {\mathcal N} (\tau , x) - {\mathcal N}_{\sst{(0)}} (x)$, $\bar{\mathcal P} (\tau , x) \equiv {\mathcal P} (\tau , x) - {\mathcal P}_{\sst{(0)}} ( x)$ tend to zero as $x \in U$ is fixed and $\tau \to +\infty$.


\section{Asymptotic geometrical structure on cosmological singularities \label{pff}} 

In the previous sections, we studied in detail the asymptotic dynamics of the gravitational field in the vicinity of a spacelike singularity in Iwasawa variables. In particular, we have seen that some of the variables have limits when $\t \rightarrow +\infty$, \ie the $Q$'s and $P$'s  while the $\b$'s have no limits. Moreover, we have argued that in  the `chaotic' case the $\pi_a$'s tend chaotically to zero. Of course these Iwasawa variables are dependent on the choice of coframe $\o^i := \o^i{}_j(x) dx^j$ used in the equation (\ref{metric}) [where $\o^i$ could be simply a coordinate coframe]. This raises the question to know whether the Iwasawa variables, despite their `gauge dependence' capture some well defined geometrical structure at the BKL limit and what is this structure. In the non--chaotic case, this question has a clear answer. Indeed, the frames that diagonalise the second fundamental form with respect to the metric have a well defined limit at the singularity. They therefore provide a fields of `directional frames' at the singularity, \ie a field of frames considered modulo rescalings of each frame vector. In this section, we will investigate the chaotic case. \newline

Our starting point is the existence of many variables having finite limits at the singularity, namely the $\cn^a{}_i$'s and $\cp^i{}_a$'s, (say $\cn^a{}_i (\tau , x) \rightarrow \cn_{\sst{(0)}}{}^a{}_i, (x)\, \cp^i{}_a(\tau, x) \rightarrow \cp_{\sst{(0)}}{}^i{}_a(x)$ as $\tau \rightarrow + \infty$). The problem is, however, that the quantities $\cn_{\sst{(0)}}{}^a{}_i$ and $\cp_{\sst{(0)}}{}^i{}_a$ do not have, a priori, a clear geometrical meaning because they depend on the coframe $\o^i$ used on $M_d$. 
One way of addressing this issue is to act on the coframe $\o^i$ with an arbitrary local transformation $\Lambda \in GL(d,\RR)$ to investigate what information is left invariant in $\cn_{\sst{(0)}}{}^a{}_i(x)$ and $\cp_{\sst{(0)}}{}^i{}_a(x)$ when `rotating' the coframe by an arbitrary $\La$. 
We will try to assign canonical values to the Iwasawa variables $\cn_{\sst{(0)}}{}^a{}_i(x)$ and $\cp_{\sst{(0)}}{}^i{}_a (x)$ at the singularity\footnote{Note that, in the non--chaotic case, we have access to more information, namely the non--zero limits of the $\pi$'s.} by means of a suitable $\La$ (e.g. in the non--chaotic case we can find $\La$'s such that: $\cn_{\sst{(0)}}{}^a{}_i(x) \stackrel{\La}{\rightarrow} \d^a{}_i$ and $\cp_{\sst{(0)}}{}^i{}_a(x) \stackrel{\La}{\rightarrow} 0$). If we are able to assign canonical values to \emph{all} asymptotic values of the $\cn_{\sst{(0)}}$'s and $\cp_{\sst{(0)}}$'s, this would mean that we would have again privileged directions at the singularity like in the non--chaotic case. We will see that the situation is actually more subtle than this.  \newline

A general matrix $\La \in GL(d,\RR)$ can be decomposed into three parts (i) a diagonal part, (ii) an upper triangular matrix (with ones on the diagonal) matrix and (iii) a lower diagonal matrix (with ones on the diagonal). \newline

(i) The action of the diagonal part of $\La$ consists in shifting the values of $\b$'s. However, as the $\b$'s have no limit as $\tau \rightarrow + \infty$ it is not clear how to extract some geometrical meaning from such shifts of the $\b$'s. \newline

(ii) Concerning the action of the upper triangular part, let us show that it can be used to fix the asymptotic values of the $\cn_{\sst{(0)}}{}^a{}_i$'s to be $\delta^a{}_i$. The crucial point is that if we act on the coframe by an \emph{upper} triangular matrix, \ie  $\o^{\prime \, i} = \La^i{}_j \o^j$, since 
\beq 
ds^2 = \sum_a e^{-2\b^a} \cn^a{}_i \cn^a{}_j \o^i \o^j
\, , \nn
\eeq
must be equal to $ds^{\prime \, 2} = \sum_a e^{-2\b^{\prime \, a}} \cn^{\prime \, a}{}_i \cn^{\prime a}{}_j \o^{\prime \, i} \o^{\prime \, j}$ and since the transformation is defined by demanding that $\cn$ and $\cn^\prime$ be both upper triangular, we easily see that 
\beq
\cn^a{}_i =   \cn^\prime{}^a{}_j \La^j{}_i \hspace{1cm} \mathrm{(for \, upper \,  triangular \, } \La \, \mathrm{ only)}. \nn
\eeq
This result is actually valid for any $\tau$, and yields, in the limit $\tau \rightarrow + \infty$, $\cn_{\sst{(0)}} =  \cn_{\sst{(0)}}^\prime \La $. Now it suffices to perform the transformation with $\Lambda^i{}_j$ chosen to be the upper triangular matrix $\Lambda^i{}_j = \cn_{\sst{(0)}}{}^{ i}{}_j$ to fix the canonical values of the $\cn_{\sst{(0)}}^\prime$'s to be the unit matrix. Note that after this fixing of the frame (such that $\cn_{\sst{(0)}}^\prime= \d$) the frame $\o^{\prime \, i}$ becomes identical at the singularity with the limiting Iwasawa frame $\th^a = \cn^{\prime \, a}{}_i \o^{\prime \, i}  = \cn^a{}_i \o^i$. \newline

(iii) Let us now consider the effect of the remaining freedom in a general $\La \in GL(d,\RR)$, \ie a \emph{lower} triangular matrix, namely,
\beq 
\Lambda = \left(
\begin{array}{ccccc} 1 & 0&0 & 0 &0\\
 \Lambda^2{}_1&1 & 0 & 0 & 0 \\
 \Lambda^3{}_1& \Lambda^3{}_2 & \cdots & 0 & 0 \\
\vdots & \vdots & \ddots & 1& 0 \\
 \Lambda^d{}_1 &  \Lambda^d{}_2 & \hdots &\Lambda^d{}_{d-1} & 1\\
\end{array}
\right) \, . \nn 
\eeq 
The action of any $\La$ (upper or lower triangular) on the frame components of the metric $ds^2 = g_{ij} \o^i \o^j = g^\prime_{ij} \o^{\prime \, i } \o^{\prime \, j}$ is always given by the following linear action, 
\beq
g^{\prime}_{ij} (\tau)&=& \Lambda^{-1 k}{}_i g_{kl}(\tau)  \Lambda^{-1 l}{}_j 
\, . \label{lowdiagchange}
\eeq
However, the \emph{induced} action of such a $\La$ on the Iwasawa variables $\b(\tau)$ and $\cn(\tau)$ parametrizing $g_{ij}(\tau)$ is somewhat complicated when $\La$ is lower diagonal because it is \emph{non linear} (contrarily to the simpler case just discussed, of an upper triangular matrix whose action was nicely compatible with the upper triangular nature of $\cn$ and was thereby linear). 
Let us then consider 
the case where the remaining lower triangular matrix $\La$ is close to the identity, say 
\beq 
\Lambda &=& 1 +  \lambda \nn \, ,\eeq
with infinitesimal (strictly lower triangular) $\lambda$. In addition, as we always assume that we have already used an upper triangular matrix $\Lambda^+$ to fix $\cn_{\sst{(0)}}$ (after dropping the primes) to the identity, we can write that $\cn(\tau)$ is of the form,
\beq
\cn(\tau )  &=& 1 + n(\tau) \, , \nn
\eeq 
with $n(\tau) \rightarrow  0$ as $\tau \rightarrow + \infty$. As both  matrices $\lambda$ and $\cn(\tau)$ can be treated as infinitesimal elements, we will neglect terms of order $O(n^2)$ and $O(\lambda^2)$. 
Note also that, as we are near a situation where $\cn^a{}_i \sim \d^a_i$, the distinction between the `a' type indices and `i' type indices disappear. When replacing the Iwasawa decomposition $g_{ij} = \sum_a e^{-2 \b^a} \cn^a{}_i \cn^a{}_j$ in equation (\ref{lowdiagchange}), we find at the linear approximation,
\beq
n^{\prime i}{}_j &=& n^i{}_j + [n,\lambda ]^i{}_j  + O(e^{-2(\b^j-\b^i)})\, , \nn \\
e^{-2\beta^{\prime i}} &=& e^{-2\beta^{i}}  (1 + 2[n, \lambda]^i{}_i) \hspace{2cm} \mathrm{no \, sum \, over \, }i \, .
\label{translower}
\eeq
As the $\cp$'s are the canonically conjugate to the $\cn=1+n$, the law of transformation of the $\cp$'s is obtained by using the `conservation' of the canonical form $\cp d \cn + \pi d\b = \cp^\prime d\cn^\prime + \pi^\prime d\b^\prime$. 
We then easily find
\beq
\cp^{i}{}_j &=& \cp^{\prime i}{}_j + [\lambda,\cp^\prime]^i{}_j - \lambda^i{}_j (\p^{\prime j} - \pi^{\prime i})+ O(e^{-2(\b^j-\b^i)}) \nn \eeq
In addition we see from equation (\ref{translower}) above that the limit $n(\tau) \rightarrow 0$ is invariant under such a lower triangular transformation $\lambda$, so that the canonical value $\cn_{\sst{(0)}}= \d^i_a$ is left fixed by such a $\La$. Since at the BKL limit, the exponential `symmetry walls' $e^{-2(\b^j- \b^i)}$ ($j>i$) vanish and that, in the chaotic case, the momentum conjugate to the $\b$'s is going to zero (after each collision, they are `redshifted' \cite{Damour:2002et}), we obtain the following action of the lower triangle $\La = 1 + \lambda$ on the limiting values of the $\cp$'s, \ie $\cp_{\sst{(0)}}$'s:
\beq
\cp_{\sst{(0)}}{}^{\prime \,  i}{}_j &=& \cp_{\sst{(0)}}{}^i{}_j + [\lambda,\cp_{\sst{(0)}}]^i{}_j \, . \label{actionlambdap} \eeq
In view of the strictly lower triangular nature of both $\lambda$ and $\cp$ it is easily seen that the transformation law (\ref{actionlambdap}) implies that the elements of $\cp_{\sst{(0)}}$ on the first lower diagonal are left invariant:
$$\cp^\prime_{\sst{(0)}}{}^{i+1}{}_i = \cp_{\sst{(0)}}{}^{i+1}{}_i \, . $$
Consistently with our general requirement of having non vanishing dominant symmetry wall coefficients, we assume that all the constants $\cp_{\sst{(0)}}{}^{i+1}{}_i$ are non vanishing.\footnote{Such a nilpotent element $\cp_{\sst{(0)}}$ is called `regular' in the mathematical literature.} Then it is easily seen that by choosing a suitable $\lambda$ one can change at will the values of the $\cp_{\sst{(0)}}$'s on the lower diagonals, $\cp_{\sst{(0)}}{}^{i+n}{}_i$ ($n\geq 2$). This proves that there exists a $\La$ such that we can fix $\cp_{\sst{(0)}}$ to the following canonical form,
\beq \cp_\circ = \left(
\begin{array}{ccccc} 0 & 0&0 & 0 &0\\
 \cp^2{}_1&0 & 0 & 0 & 0 \\
0 & \cp^3{}_2 & \cdots & 0 & 0 \\
0 & 0 & \ddots & 0 & 0 \\
0 & 0 & \hdots &\cp^d{}_{d-1} & 0 \\
\end{array} \right) \, .
\label{fixp}
\eeq
Let us now study what are the coframe changes that leaves the canonical  of $\cn_{\sst{(0)}}$ ($\cn_{\sst{(0)}} = \d$) and $\cp_{\sst{(0)}}$ ($\cp_{\sst{(0)}}=\cp_\circ$) invariant. We already know that the canonical form $\cn_{\sst{(0)}}=\d$  fixes the upper part of $\La$ to be unity. As for the lower diagonal part $\La = 1 + \lambda$, the request that $\cp_{\sst{(0)}}$ be fixed to $\cp_\circ$ implies, from the equation (\ref{actionlambdap}), the condition
\beq
\, [\lambda,\cp_\circ]^i{}_j = 0 \, .
\eeq
To analyze the consequences of this constraint, we decompose $\l$ into a sum of matrices with non vanishing elements only on one of its `lower' diagonals, \ie $\lambda =\lambda_1 + \lambda_2 + ...+\lambda_{d-1}$ where the only non zero elements of $\lambda_n$ are $(\lambda_n)^{i+n}{}_i$. 
It is then easily seen that $\lambda_1$ must be proportional to $\cp_\circ$. Then one similarly finds that $\lambda_2 \propto \cp_\circ^2$ etc... Finally the most general $\La = 1 + \lambda$ fixing $\cp_{\sst{(0)}}$ to its canonical value $\cp_\circ$ is found to be of the form,
\beq 
\lambda = \alpha_1 \cp_\circ + \a_2 \cp_\circ^2 + \a_3 \cp_\circ^3 + ... + \a_{d-1} \cp_\circ^{d-1} \label{stab}
\eeq
for some constants $\a_n$.  \newline

Let us discuss the geometrical meaning of our findings. We consider again a general $\La \in GL(d,\RR)$, containing both upper and lower triangular parts (for the reasons explained above we do not consider the diagonal part). If we had been able to define a canonical form whose `stabiliser' in $GL(d,\RR)$ had been only the unit matrix, this would have meant the existence of a preferred \emph{directional frame} (frame modulo rescalings) at each spatial point $x$ `on the singularity'. On the other hand, if the stabilizer had been the full group $GL(d,\RR)$, this would have meant that no preferred frame at all remained on the singularity. 
Actually we have a intermediate situation, our stabiliser $S$ is a proper subgroup of $GL(d,\RR)$, $\{\mathun \} \subset S \subset GL(d,\RR)$. It defines an equivalence class of directional frames that we can call a \emph{partially framed flag}.\footnote{Let us recall that a (complete) \emph{flag} can be seen as the equivalence class of `directional frames' with the following equivalence relations. A `directional frame' given by the directions $\{v_1,...,v_d\}$ is equivalent to the `directional frames' $\{v_1^\prime,...,v^\prime_d\}$ constructed by picking a first direction along the vector $v_1$ ($v^\prime_1 \propto v_1$), then a second direction $v_2^\prime$ belonging to 2--plane spanned by $v_1$ and $v_2$ ($v_2^\prime \propto v_2 + \a v_1$), a third direction belonging to the 3--plane spanned by $v_1$, $v_2$ and $v_3$ ($v_3^\prime \propto v_3 + \b v_2 + \g v_1$), and so on up to a last direction along a vector $v^\prime_n$ which is an arbitrary vector in $\RR^d$. In other words,
the stabiliser of a flag is the full subgroup of lower triangular matrices of $GL(d,\RR)$. } 
The elements of our stabiliser $S$ are given at the infinitesimal level by the formula (\ref{stab}). Therefore, an element $s$ of $S$
can be written as
\beq
s = e^{\alpha_1 \cp_\circ + \a_2 \cp_\circ^2 + \a_3 \cp_\circ^3 + ... + \a_{d-1} \cp_\circ^{d-1} }\label{Stab}
\eeq
where $\a_1, \, \a_2, \, \a_3, ... , \, \a_{d-1}$ are constants. The element $s$ of $S$ acts on the coframe $\o^i$ as,
\beq
\o^{\prime \, i } = s^i{}_j \o^j \label{equivclasscoframe} \, .
\eeq 
It is then easily checked that $S$ is a commutative group of dimension $d-1$. More precisely, in view of the fact that the various powers of the matrix $\cp_\circ$ commute among themselves, one finds that the group composition of two elements of $S$ is simply given by 
\beq
s(\a_1,\a_2,\dots, \a_{d-1}) \circ s(\a^\prime_1,\a^\prime_2,\dots, \a^\prime_{d-1})=s(\a_1+\a^\prime_1,\a_2+\a^\prime_2,\dots, \a_{d-1}+ \a^\prime_{d-1})
\, . 
\eeq 
[In mathematical terminology, $S$ is a unipotent abelian subgroup of the Borel subgroup of $GL(d,\RR)$.] Explicitly, the 
matrix elements of $s$ (\ref{Stab}) read as follow,
\beq
s^n{}_i = \sum_{j \in (1,\dots,d-1) \arrowvert \exists m \in \{ \NN^\star \arrowvert mj = n- i\}}
{\a_j^m \over m!} \cp_{\sst{(0)}}{}^n{}_{n-1} \dots \cp_{\sst{(0)}}{}^{i+1}{}_{i} \, .
\eeq
For instance in $d=3$, the matrix $s$ reads,
\beq 
s = \left(
\begin{array}{ccc}
1 & 0 & 0 \\
\a_1  \cp_{\sst{(0)}}{}^2{}_{1} & 1 & 0 \\
({1\over 2} \a_1^2 + \a_2)\cp_{\sst{(0)}}{}^3{}_{2}\cp_{\sst{(0)}}{}^2{}_{1} & \a_1 \cp_{\sst{(0)}}{}^3{}_{2} &1
\end{array} 
\right)
\, . 
\eeq
Therefore, in $d=3$, the class (defined by the relation (\ref{equivclasscoframe})) of coframes equivalent to some given coframe $\o^{_\prime \, i}$ is explicitly given by: 
\beq
\o^{\prime \, 1 } &=& \o^1 \nn  \\
\o^{\prime \, 2 } &=& \o^2 +  \a_1  \cp_{\sst{(0)}}{}^2{}_{1} \, \o^1 \nn \\
\o^{\prime \, 3 } &=& \o^3 +  \a_1  \cp_{\sst{(0)}}{}^3{}_{2} \, \o^2+({1\over 2} \a_1^2 + \a_2)\cp_{\sst{(0)}}{}^3{}_{2}\cp_{\sst{(0)}}{}^2{}_{1} \o^1
\, . \label{3dcoframeclass} 
\eeq
By duality between frames and coframes ($<\o^i, e_j> = \d^i_j$), one can then easily deduce the corresponding equivalence classes of frames. The equivalence class of a frame $e_i$ is given by
\beq
[e_i ] = \{e_i^\prime =  e_j s^{-1}{}^j{}_i\arrowvert s \in S \} \, . \label{equivclassframe} 
\eeq
For $d=3$, we have explicitly, 
\beq
e^\prime_1& =&  e_1 -\a_1  \cp_{\sst{(0)}}{}^2{}_{1}  e_2 +({1\over 2} \a_1^2 - \a_2)\cp_{\sst{(0)}}{}^3{}_{2}\cp_{\sst{(0)}}{}^2{}_{1} e_3 \nn \\
e^\prime_2& =& e_2 -  \a_1  \cp_{\sst{(0)}}{}^3{}_{2} e_3 \nn \\
e^\prime_3 &=& e_3 \, . \label{3dframeclass}
\eeq

\emph{Summary}: The equivalence class of frames and coframes with respect to which the limiting values of $\cn_{\sst{(0)}}$ and $\cp_{\sst{(0)}}$ take the canonical values $\cn_{\sst{(0)}}= \d$ and $\cp_{\sst{(0)}}=\cp_\circ$ (\ref{fixp}) is described by the relations (\ref{equivclassframe}) and (\ref{equivclasscoframe}) which contain $d-1$ arbitrary parameters. This equivalence class defines a privileged geometrical structure at the singularity, which we call a \emph{partially framed flag}. For instance, in $d=3$ this partially framed flag comprises: (i) one privileged direction $e_3$, which is independent of the basic frame $\o^i$, (ii) then the equivalence class of the vector $e_2$ must lie in a privileged 2--plane containing $e_3$, (iii) finally, after having chosen a representative $e_2$ in the class $[e_2]$ (\ie $\a_1$ is fixed), the third vector $e_1$ must lie in a privileged 2--plane defined by $ e_1 -\a_1  \cp_{\sst{(0)}}{}^2{}_{1}  e_2 +{1\over 2} \a_1^2\cp_{\sst{(0)}}{}^3{}_{2}\cp_{\sst{(0)}}{}^2{}_{1} e_3$ and $e_3$. [This is different from a flag where $e_1$ would have been an arbitrary direction in the full $\RR^3$.]\newline

It is remarkable that at the `chaotic' BKL limit, we have a such geometrical structure left. At the singularity, we could have guessed that nothing is left from the metric structure because of the chaotic character of the asymptotic dynamics. Let us note that our results have been partially anticipated in \cite{BKL2} where a law of `rotation of Kasner axes' was derived. This law is somewhat similar to our results (\ref{3dcoframeclass}) but, actually, it has a different physical meaning. Indeed, the first Kasner axis $l$ which is preserved in the law approximatively derived in \cite{BKL2} is supposed to belong to the `growing' eigen axis $p_l = p_1 < 0$, so that it would be a different axis that would be preserved during further collisions. One would then have no privileged direction at the singularity.
As we have shown here, there exists a well defined geometrical structure at the singularity: a \emph{partially framed flag} which is rather `rigid' in the sense that it depends only on $d-1$ arbitrary parameters (while a generic flag would involve $d(d-1)/2$ arbitrary parameters). 

\section{Conclusion}

In this paper, we started by reconsidering the asymptotic dynamics, in the vicinity of a spacelike singularity, of the fields for `non-chaotic' Einstein--matter systems.  We have outlined a new proof that gives this asymptotic dynamics (which is essentially given, at each spatial point, by a monotone power--law solution in terms of the proper time). Our method is based on the Iwasawa decomposition of the spatial metric and on the Hamiltonian formulation of Einstein--matter systems.  As in references \cite{Andersson:2000cv,Damour:2002tc}, we have used the Fuchs theorem to conclude. More precisely, we have defined an asymptotic system of equations which is a system of ODEs and we have also defined asymptotic constraints.  Next, we have shown that the `differenced variables' (\ie the differences between the solution of the exact Einstein--matter constrained equations and the solution of the asymptotic Einstein--matter constrained equations) obey a Fuchsian system.  A solution of the constrained asymptotic system, together with initial data, can thus be used to parametrize an exact constrained solution. The advantages of our formulation is that it is shorter, more transparent (the neglected terms are walls / subdominant walls) and that we avoid the problem of the symmetry of the metric encountered in \cite{Andersson:2000cv,Damour:2002tc}. In appendix \ref{appB}, we discuss the spatial domain on which a Fuchsian analysis can be applied in this context and point out that our method cannot be used in some zero--measure co--dimension 2 submanifolds (in reference \cite{Andersson:2000cv,Damour:2002tc}, there is a quite involved construction to deal with these submanifolds).  We also showed that this problem originates in the `spinorial' nature of the eigenvectors of the second fundamental form around the submanifold where 2 (or more) eigenvalues coincide. \newline

We next turned to the `chaotic' Einstein--matter systems and formulated a precise statement for the chaotic BKL behavior. This is achieved along the same lines as our formulation of the non--chaotic case. We parametrize, at each spatial point, the generic solution of the asymptotic behavior of the fields close to a spacelike singularity in terms of a constrained system of ODEs (and some initial data).  Then we argue that the difference between the solution of the exact `chaotic' constrained Einstein--matter system and the asymptotic system just defined satisfies a 'generalized Fuchs system'. 
We leave to others the task of proving that such `generalized Fuchsian systems' admit a unique, asymptotical vanishing, solution. Our purpose here was  mainly to formulate, in precise mathematical and physical terms, this asymptotic characterization of `chaotic' solutions of Einstein--matter systems. \newline

Finally,  we addressed the question of the existence of some asymptotic geometrical structure defined \emph{at }Êthe singularity for a chaotic system. We knew that some of the metric variables had finite limits at the singularity and it was therefore natural to wonder whether we could extract some geometrical structure from these limiting values.  A first slight, we could expect that the chaotic nature of the asymptotic dynamics would destroy any structure at the singularity. We showed that it is not the case: partially framed flags can be defined at the singularity.  These partially framed flags are, as their name indicate, more `rigid' than flags and less `rigid' than frames.

\section*{Acknowledgements}

It is a pleasure to thank Yvonne Choquet--Bruhat for instructive exchanges about reference \cite{choquet-bruhat} and her improvement of the Fuchsian theorems that we have used.  We would also like to thank Offer Gabber, Jim Isenberg, Maxime Kontsevich, Laurent Lafforgue and Vincent Moncrief for interesting discussions. SdB thanks Daniel Persson and St\'ephane Detournay for discussions. Ryan Budney is acknowledged for suggesting the name \emph{partially framed flag}.  The work of TD was partially supported by ICRANet, Pescara, Italy. The work of SdB was supported by the RTN project ``ForcesUniverse'' MRTN-CT-2004-005104 during the elaboration of this paper.


\appendix

\section{Evolution equations of the Hamiltonian and momentum constraints \label{hamform}}

The purpose of this appendix is to obtain the evolution equations for the exact Hamiltonian constraint $\ch$ (\ref{hamconst}) and exact momentum constraints (\ref{momconst}) in our gauge choices. Here we do not consider matter for simplicity and we work in a coordinate basis. These evolution equations can be derived from the Bianchi identities and the evolution equations for the spatial metric. In this perspective, we have to know what are the evolution equations in the gauge we are interested in. \newline

The first order action of pure gravity in $D=d+1$ spacetime dimensions $S[g_{ij},\pi^{ij},\tilde N,N^i]$ reads,
\beq
S[g_{ij},\pi^{ij},\tilde N,N^i]= \int dx^0 d^d x ( \dot{g}_{ij} \pi^{ij} -\tilde N \ch - N^i \ch_i ) \, .
\eeq 
Its variation with respect to $\pi^{ij}$ can be understood as the definition of the $\pi_{ij}$ in terms of $\dot g_{ij}$, the variations with respect to $g_{ij}, \, \tilde N, \, N^i$ giving respectively the equations of motion, the Hamiltonian constraint and the momentum constraints.\footnote{We recall the following relationships, \newline
\begin{tabular}{p{6cm}p{6cm}}
\beq ^{\sst{(D)}} g_{\mu \n}Ê= 
\left(\begin{array}{cc}
N_k N^k-\tilde N^2 g & N_j \\
N_i & g_{ij} 
\end{array}Ê\right) \nn \eeq
& 
\beq ^{\sst{(D)}} g^{\mu \n}Ê= 
\left(\begin{array}{cc}
-{1 \over \tilde N^2 g} & {N^j  \over \tilde N^2 g} \\
 {N^i \over \tilde N^2 g} & g^{ij} -{ N^i N^j \over \tilde N^2 g} 
\end{array}Ê\right) \, , \nn \eeq
\end{tabular} \newline 
where $\mu = (0,i)$ and $\nu=(0,j)$.} 
Note that the dot means a derivation with respect to $x^0$. Let us now use the Einstein--Hilbert action, $ \int d^D x \, \sqrt{- ^{\sst{(D)}}g} ^{\sst{(D)}}R$ in the so--called Palatini formalism (\ie the Christoffel symbols $\G^\m_{\n\rho}$ are considered to be independent of the metric elements $g_{\m\n}$) to determine the link between the Hamiltonian equations of motion $\d S / \d g_{ij} =0$ and the usual Einstein equations $^{\sst{(D)}}G_{\m\n} = 0$. 
The variation of the Einstein--Hilbert--Palatini action gives,
\beq
\d S_{EHP} &=& \int d^Dx \sqrt{- ^{\sst{(D)}}g} ( -{}^{\sst{(D)}}G^{\mu \nu} \d g_{\m\n} + g^{\m\n}(\d \G^\lambda{}_{\m\n;\lambda} - \d \G^{ \lambda}_{\m\lambda;\n}) \nn \\
&=& \int d^Dx ( -\sqrt{- ^{\sst{(D)}}g} ^{\sst{(D)}}G^{\mu \nu} \d g_{\m\n} - ((\sqrt{-g} g^{\m\n})_{;\rho} - \d^\n_\rho (g^{\m \lambda}\sqrt{-g})_{;\lambda}) \d \G^\rho_{\m\n}) \, , 
\label{varaction}
\eeq
where $ ^{\sst{(D)}}g$ denotes the determinant of the spactime metric, $ ^{\sst{(D)}}G_{\m\n}$ is the Einstein tensor and the second equality is obtained by integration by parts (we neglect the boundary terms). For simplicity, let us assume that $N^i=0$ on--shell (but keeping $\d N^i \neq 0$), and that most  of the usual relations between $\G^\m{}_{\n\rho}$ and the derivatives of $g_{\m\n}$ are constrained to hold, namely 
\beq
\G^0{}_{00} &=& {1\over 2} {g_{,0} \over g}+{\tilde N_{,0} \over \tilde N} \nn \\
\G^0{}_{0i} &=&  \G^0{}_{i0} ={1\over 2} { g_{,i} \over g} +{\tilde N_{,i} \over \tilde N}\nn \\
\G^i{}_{00} &=&{1\over 2} \tilde N^2 g^{ij} g_{,j} +g^{ij} \tilde N \tilde N_{,j} g \nn \\
\G^i{}_{jk} &=& {1\over 2} g^{il}(g_{lj,k} + g_{lk,j} - g_{jk,l}) \, .
\eeq 
Moreover we require that $\G^0{}_{jk}$ is related to the $\G^j{}_{k0} = \G^j{}_{0k}$ via $\G^j{}_{0k} = g g^{ji} \G^0{}_{ik}$. We can then verify that the terms in $\d \G$ can be ignored to compute the functional derivative of $S$ with respect to $g_{ij}, \, \tilde N $ and $N^i$ (the coefficient in front of these variation in (\ref{varaction}) vanish). 
It is then straightforward to derive the following relations,
\beq
{\d S_{EHP} \over \d g_{ij}}  
&=& \tilde N g ( -^{\sst{(D)}}G^{ij} + \tilde N^2 g  {}^{\sst{(D)}}G^{00}   g^{ij}) + O(N^k) \\
{\d S_{EHP} \over \d \tilde N} &=& 2  ^{\sst{(D)}}G^{00} \tilde N^2 g^2 + O(N^k) \\
{\d S_{EHP} \over \d N^i} &=& -{2 \tilde N g} ^{\sst{(D)}}G^{i0}+ O(N^k) \, .
\eeq
Note that we did not write explicitly the terms proportional to $N_i$ because we will work in the gauge $N_i=0$. On the other hand, we have,
\beq
{\d S \over \d g_{ij}} &=& - \dot \pi^{ij} - \tilde N {\d \ch \over \d g_{ij}} + O(N^k) \\
{\d S \over \d \tilde N} &=& -\ch \\
{\d S \over \d N_i} &=&- \ch^i
\eeq
When identifying $\d S_{EHP}$ with $\d S$, one obtains,
\beq
\ch = -2  g^2 \tilde N^2 \,  \, {}^{\sst{(D)}}G^{00} + O(N^k)= -{2\over \tilde N^2} G_{00} + O(N^k)\, , \label{ch}\\
\ch^i = 2 { g} \tilde N \, \,^{\sst{(D)}}G^{i0} + O(N^k)= - {2 \over \tilde N} g^{ij}G_{0j} + O(N^k)\, . \label{chi} 
\eeq
On the other hand, the equations of motion $\d S /\d g_{ij}=0$ are found to be equivalent to the equations
\beq
^{\sst{(D)}}G^{ij} = {1 \over 2g}  \ch   g^{ij} + O(N^k) \, . \label{eom}
\eeq
Note that our result (\ref{eom}) is linked to our choice of `rescaled lapse' $\tilde N$ as basic lapse variable. The result would be different if we were using the usual lapse $N$. \newline

Let us now use the Bianchi identities $B_\m :=\nabla_\n {}^{\sst{(D)}} G^{\n}{}_{\m} \equiv 0$ to derive the evolution equations for the constraints. We use the equality, 
\beq
B_\m= {\partial_\n (\sqrt{-{}^{\sst{(D)}}g} {}^{\sst{(D)}}G^\nu{}_\mu) \over 
\sqrt{-{}^{\sst{(D)}}g} } -{ 1 \over 2} \pa_{\m}g_{\a\b}  {}^{\sst{(D)}}G^{\a\b} \,  .\label{bianchi}
\eeq
When inserting in the expression of $B_0$ the relationship between $\ch$, $\ch_i$ and the components of the Einstein tensor (\ref{ch}, \ref{chi}) as well as the equations of motion (\ref{eom}), we obtain the evolution equation for $\ch$,
\beq
{1 \over 2 g}  \partial_\tau \ch - {\tilde N \over 2}Ê\nabla^i \ch_i - (\nabla_i \tilde N  ) \ch^i  + O(N^k) =0\, . \label{evoeqch}
\eeq
Note that $\ch_i = g_{ij} \ch^j$ and that it is a tensorial density of weight 1 while $\tilde N$ is a scalar density of weight --1. The covariant derivatives $\nabla^i \ch_i$ and $\nabla_i \tilde N$ must take into account these weights: for instance $\nabla^i \ch_i = \nabla_i \ch^i = \partial_i (g^{ij} \ch_j)$ (in a coordinate frame) and $\nabla_i \tilde N = \pa_i(\tilde N \sqrt{g})/ \sqrt{g}$. From $B_i=0$, we get
\beq
{1 \over 2g} ({\partial_\tau \ch_i \over \tilde N} - \nabla_i \ch) + O(N^k)= 0
\, . \label{eveqchi}
\eeq
Note that $\ch$ is a scalar density of weight 2 so that $\nabla_i \ch= g \pa_i(\ch / g)$. 
In our gauge choices, \ie $\tilde N = 1$ and $N^i = 0$, the equations (\ref{evoeqch}) and (\ref{eveqchi}) read, 
\beq
 \partial_\tau \ch &=& Êg\nabla^i \ch_i + g_{,i} \ch^i \, , \label{evoeqchgf} \\
 \partial_\tau \ch_i - \nabla_i \ch &=& 0
\, . \label{eveqchigf}
\eeq

\section{Fuchsian Systems \label{appFuchs}}

\subsection{Fuchs Theorem}
The general form of a Fuchsian
system (\cite{baouendi77, Kichenassamy:1997bt, choquet-bruhat} and references therein) for a vector--valued unknown function $u(t,x) = (u_1(t,x), ...,u_k(t,x))$, 
defined on an open subset of $\RR\times\RR^n$ with values in $\RR^k$, 
is
\begin{equation}
\label{fuchs0}
t \, \partial_t u + \ca(x) \, u = t^{\mu}f(t,x,u,\partial_x u),
\end{equation}
where $\partial_x u$ denotes a finite number of derivatives
of $u$ with respect to the variables $x$ (they are not restricted to be of first order); 
the function $f$ is defined on $(0,T_0]\times U_1\times U_2$  (where $U_1$ is an open subset of $\RR^n$ and $U_2$ is an open subset of $\RR^{k+nk}$) and takes values in $\RR^k$; and $\m >0$. 
$\ca$ is an analytic $k \times k$ matrix-valued function defined on $U_1$. The system (\ref{fuchs0}) is said to be \emph{Fuchsian }if the matrix $\ca(x)$ and the function $f$ fulfill 
the following conditions, 
\begin{itemize}
\item \emph{condition on $\ca$ :} the matrix  ${\cal A}(x)$ is required to
satisfy some lower boundedness condition. One sufficient condition, that has been used in several works \cite{Andersson:2000cv,Damour:2002tc}, is that
there is a constant $\epsilon$ such that Real$(\lambda) \, > \, \epsilon>0$, for
each eigenvalue $\lambda$ of $\ca$ at any point. Recently, this condition has been relaxed to requiring that there exists an $0\leq \a \leq \m$ such that 
$|\sigma|^{{\cal A}} \sigma^\alpha $ be bounded for $\sigma$ varying in the interval $0 \leq \sigma \leq 1 $ \cite{choquet-bruhat}. Essentially, this condition means that the real part of the eigenvalues of $\ca$ are everywhere strictly larger than $-\m$. This is of particular interest for us, since the matrix ${\cal A}$ relevant in our case is nilpotent so that $\arrowvert \sigma \arrowvert^\ca$ grows like a power of $\log{\sigma}$ as $\sigma \rightarrow 0$ (so that we can simply use any $\a$ in the interval $0 <\a < \mu$). 
\item \emph{condition on $f$ :} 
 the `source term' $f$ (after having factored $t^\mu$) is required
to be `regular', \ie  $f$ must possess an analytic continuation in $x$, $u$ and $\pa_x u$ and, as a function of $t$, must be continuous on $[0,t_\circ]$ for some finite time $t_\circ$.  
\end{itemize} 
For more precise conditions on $\ca$ and $f$, we refer to \cite{Andersson:2000cv} and references therein. \newline

\emph{Fuchs Theorem: } If the system (\ref{fuchs0}) satisfies the above conditions to be Fuchsian, then it 
possesses a unique solution $u$
that vanishes as $t \rightarrow + 0$ 
(see {\it e.g} \cite{Andersson:2000cv,choquet-bruhat}).
\newline

\underline{Note} : After the change of variable $t=e^{-\tau}$ (such that the singularity is now located at $\tau \rightarrow +\infty$), a Fuchsian system reads,
\begin{equation}
\label{fuchstautau}
 \, \partial_\tau u - \ca(x) \, u = e^{-\mu \tau} \bar f(\tau,x,u,\partial_x u),
\end{equation}
where, essentially, $\bar f$ must be analytic in $x, \, u , \, \pa_x u$ and bounded in $\tau $ as $\tau \rightarrow +\infty$. This is the form we shall use in the text.\newline

\subsection{Shift of the eigenvalues of ${\cal A}$}

One may wonder if it is possible to have a more precise description of how fast the solution $u$ of the Fuchsian system (\ref{fuchs0}) goes to zero when $t$ goes to zero. To answer this question, let us rewrite the system (\ref{fuchs0}) in terms of the variable $\bar u$ defined as follows,
\beq 
\bar u = t^{-\lambda }u\, , \hspace{1cm} 0<\lambda<\mu . \label{newu} \eeq
Inserting (\ref{newu}) in the system (\ref{fuchs0}), it is straightforward to obtain
\begin{equation}
\label{fuchs1}
t \, \partial_t \bar u + \bar \ca(x) \, u = t^{\bar \mu } \bar f(t,x, \bar u,  \partial_x \bar u) \, ,
\end{equation}
where
\beq 
\bar \ca &=& \ca + \lambda \mathun \, , \hspace{.7cm} \bar \mu = \mu - \lambda \nn \\
\bar f &=&  f(t,x, t^\lambda u, t^\lambda \partial_x u) \, .\nn 
\eeq
If $f$ is regular, then $\bar f$ is regular. Note that the eigenvalues of $\bar \ca$ have been shifted by $\lambda >0$ compared to those of $\ca$, so that if $\ca$ satisfies the lower boundedness conditions eigenvalues($\ca$)$>-\m$, so does $\bar \ca$ (with the corresponding $\bar \m = \m - \lambda$). Therefore the `shifted' system (\ref{fuchs1}) is again Fuchsian and we know that it admits a unique solution  $\bar u $ that vanishes when $t \rightarrow +0$. This tells us that the unique solution of the Fuchsian system (\ref{fuchs0}) that vanishes as $t \rightarrow + 0$ actually vanishes as $u= t^\lambda \bar u$ with $\bar u = o(1)$, \ie as $o(t^\lambda)$ for any $\lambda< \mu$. \newline

\emph{Summary :} the `shift' of $u$ allows us to gain a more precise information about how the unique (asymptotically vanishing) solution $u$ of the system (\ref{fuchs0}) decays as $t\rightarrow +0$. If the r.h.s. of the Fuchsian system (\ref{fuchs0}) decays as $t^\m f$
then $u$ is a  $O(t^{\mu- \epsilon})$ for any $\epsilon > 0$. When using the $\tau$ variable, this essentially means that a source term decaying as $e^{-\m \tau}$ corresponds to a unique, asymptotically vanishing, solution decaying as $e^{-\mu_-\tau}$ for any $0<\mu_-<\mu$.\newline

\underline{Note} : For completeness, let us define the notations $O$ and $o$. A function $F(t,x,p)$
defined on $(0,T_0] \times U_1 \times U_2$, where $U_1, U_2$ are open subsets of $\RR^n$ and $\RR^N$ respectively, is said to be $O(G(t))$ if
there is a constant $C$ such
that
$$
|F(t,x,p)| \leq C|G(t)| \quad \text{for $t \in (0,t_0]$, $(x,p)\in K$}.
$$
 The notation
$F=o(G(t))$ is used to indicate that $F/G$ tends to zero uniformly
on compact subsets of $U_1\times U_2$ as $t\to 0$.   \newline

\section[Some subtleties occurring in the non--chaotic case]{Subtleties occurring when some of the eigenvalues of the second fundamental form coincide in the non--chaotic case \label{appB}}

The usual AVTD approach uses a rather complicated construction to deal with the neighborhoods of points where some eigenvalues of the second fundamental form $k_{ij}$ coincide \cite{Andersson:2000cv,Damour:2002tc}. Such a complication is needed because the frame vectors that diagonalise $k_{ij}$ with respect to $g_{ij}$ are \emph{not analytic } in $x$ near such points. Here we consider the behavior of Iwasawa variables in these regions. For a full comparison one should carefully analyze the different slicing hypersurfaces in the two approaches: gaussian slicing $N=1$ in AVTD vs pseudo--gaussian slicing $\tilde N=1$ in our case.\newline

As a simple example, let us consider gravity in $D=4$ coupled to a dilaton. Let us consider for simplicity the (generic) case where two of the eigenvalues coincide on some submanifold. We choose as one of the frame vectors, the (analytic) eigenvector $e_3$ corresponding to the third eigenvalue (which is supposed to stay away from the other two). The two other analytic frame vectors are chosen to be orthogonal to $e_3$ and to each other (they are linear combinations of the eigenvectors corresponding to the nearly degenerate eigenvalue). In this orthonormal basis (or dreibein), the coefficients of the metric are $g_{ab} = \d_{ab}$, while the coefficient of the second fundamental form are given by a matrix $K$ which is of the following form,
\beq 
K = \left( 
\begin{array}{ccc}
a+c & b & 0 \\
b & -a+c & 0 \\
0 & 0 & d \end{array}
\right)
\nn \, ,\eeq
where $a, \, b, \, c$ and $d$ depend analytically on the spatial coordinates. 
The eigenvalues of $K$ are $c(x)\pm \sqrt{a(x)^2 + b(x)^2}$ and $d(x)$. Therefore, two eigenvalues will coincide when $a(x)^2 +b(x)^2$ vanishes, which means that \emph{both} $a(x)$ and $b(x)$ must vanish. This happens generically on a line in the three dimensional space since it gives us two conditions $a(x^i)=0$ and $b(x^i)=0$. 
If we were in $d$ spatial dimensions, the submanifold $\cl$ where two eigenvalues coincide would again be defined by the vanishing of some $a(x)^2+b(x)^2$ and therefore be a codimension 2 submanifold.
For convenience, let us replace the quantities $a(x^i)$ and $b(x^i)$ by $\rho(x^i)$ and $\th(x^i)$ such that  $a(x^i) = \rho(x^i) \cos{\th(x^i)}$ and $ b(x^i) = \rho(x^i) \sin{\th(x^i)}$. Since $a$ and $b$ have generically their values between -$\infty $ and $+\infty$, we have $\rho \, \in \, [ 0, \infty ]$ and $ \theta \in \, [0, 2 \pi [$. 
Let us consider the Kasner metric, expressed in the time $\tau= -\ln{t}$, $g_{\sst{[0]}}(\tau,x^i) = e^{-K\tau}$
(see  \cite{Andersson:2000cv,Damour:2002tc}). We have
\beq
g_{\sst{[0]}} = \left( 
\begin{array}{ccc}
e^{c\tau}(\cosh{\rho \tau} +\cos{\th}  \sinh{\rho \tau}) & e^{c\tau} \sin{\th} \sinh{\rho\tau}& 0 \\
e^{c\tau}\sin{\th} \sinh{\rho\tau} & e^{c\tau}(\cosh{\rho \tau}- \cos{\th} \sinh{\rho \tau} ) & 0 \\
0 & 0 &  e^{d \tau} \end{array}
\right)
\label{kasnercoinc}  \, .
\eeq
To compute the Iwasawa variables corresponding to the metric (\ref{kasnercoinc}), we use the following explicit formulas (\ref{iwaexplicit}),
\beq
\b^1&=& -{1\over 2} c \tau - {1\over 2} \ln{({1 +\cos{\th} \over 2} e^{\rho \tau}+{1 -\cos{\th} \over 2} e^{-\rho \tau} )} \nn \\
\b^2&=& -{1\over 2} c \tau + {1\over 2} \ln{({1 +\cos{\th} \over 2} e^{\rho \tau}+{1 -\cos{\th} \over 2} e^{-\rho \tau} )} \nn \\
\b^3 &=& d \tau \nn  \\
\cn^1{}_2 
&=&{\sin{\th} (e^{\rho \tau} - e^{-\rho \tau})\over (1+ \cos{\th} )e^{\rho \tau} + (1- \cos{\th} )e^{-\rho \tau} } \nn \\
\cn^1{}_3 &=& 0   \nn \\
\cn^2{}_3 &=& 0  \label{kasneriwa}
\eeq
The crucial point is that the co--dimension 1 submanifold $\Sigma$ defined by the equation $1+\cos{\th}=0$ plays a singular role in the formulas (\ref{kasneriwa}). Indeed, the exponential growing term $e^{\rho \tau}$ always appears in the combination $(1+\cos{\th}) e^{\rho \tau}$. Therefore, in the open domain $1+\cos{\th}\neq 0$ and $\rho = + \sqrt{a^2 +b^2} >0$, we have the generic Iwasawa behavior that the $\cn$'s have a finite limit as $\tau \rightarrow +\infty$, namely 
\beq
\cn^1{}_{2 \, \sst{(0)}} = \lim_{\tau \rightarrow + \infty} \cn^1{}_{2}(\tau) = { \sin{\th} \over 1 + \cos{\th}} \, \nn \eeq
while the $\b$'s  have the following asymptotic behavior,
\beq 
\b^1 \sim - {1 \over 2} (c+\rho) \tau - {1 \over 2} \ln{({1 +\cos{\th} \over 2})} \, , \nn \\
\b^2 \sim - {1 \over 2} (c-\rho) \tau + {1 \over 2} \ln{({1 +\cos{\th} \over 2})} \, , \nn \\
\b^3  = d\tau  \, . 
\eeq
Note also that, because $\rho >0$, we have the usual asymptotic ordering $\b^1 \leq \b^2$. However, we see that the asymptotic limit $\cn^1{}_{2 \, \sst{(0)} }$, which depends only on spatial variables, becomes \emph{singular} on $\Sigma$ (codimension 1 submanifold (with boundary) where $\cos{\th} = -1$) which is, in our $d=3$ case, a half--membrane ending on the line $a=b=0$, where the eigenvalues of $K$ coincide. More precisely, $\cn^1{}_{2 \, \sst{(0)} }$ tends to $+\infty$ as $\th \rightarrow \pi^+$ and tends to $- \infty$ as $\th \rightarrow \pi^-$. Correlatively the behavior of the $\b$'s become singular on $\Sigma$. We have the asymptotic behavior $\b^a (\tau,x^i) \sim p_\circ^a(x^i) \tau + \b^a_\circ(x^i) $ where, e.g. $\b^1_\circ(x^i) = -{1 \over 2} \ln{({1 + \cos{\th} \over 2})}$ and $\b^2_\circ(x^i) = {1 \over 2} \ln{({1 + \cos{\th} \over 2})}$ both become singular on $\Sigma$, while $p_\circ^1= -{1\over 2} (c+\rho)$, $p_\circ^2 =-{1\over 2}(c-\rho)$ [we recall that $\rho(x^i) = +\sqrt{a^2(x^i) + b^2(x^i)}$]. Note also that, if one sits on $\Sigma$, one has the asymptotic behavior $\cn^1{}_2(\tau,x^i)=0$ and $\b^1(\tau,x^i)= -{1 \over 2}(c- \rho) \tau, \, \b^2(\tau,x^i) = -{1 \over 2}(c+ \rho) \tau$ where the signs of the $\rho$ terms in $\b^1$ and $\b^2$ are exchanged compared to the asymptotic behavior outside of $\Sigma$. In particular, on $\Sigma$ we have, asymptotically, $\b^1(\tau,x^i) > \b^2(\tau,x^i) $, which contrasts with the generic result that asymptotically $\b^1 \leq \b^2$. This unusual behavior is the sign that the coefficient of the $e^{-2 (\b^2-\b^1)}$ symmetry wall \emph{vanishes} on $\Sigma$. As said above, in our treatment we neglected this possibility on the account that it is non generic (as the coefficient in question is a square). We see now that this non generic behavior necessarily occurs on some codimension 1 submanifolds ending on the codimension 2 submanifolds where 2 eigenvalues of $K$ coincide. \newline

However, let us emphasize that the location of the singular codimension 1 submanifold $\Sigma$ is not geometrically fixed, but is somewhat arbitrary apart from the fact that it necessarily ends on the codimension 2 submanifold $\cl$ where 2 eigenvalues coincide. Indeed, let us show that, by using a suitable, $x$--dependent local $SO(2,\RR)$ transformation, one can move $\Sigma$ around $\cl$, in a manner similar to an ordinary--life flag moving around its pole. More precisely, let us perform the following (spatially dependent) rotation of the first  two vectors $e_1 $ and $e_2$ of our orthonormal frame,
\beq
\left( 
\begin{array}{c} 
e_1^\prime \\
e_2^\prime  \\
e_3^\prime 
\end{array} \right) = 
\left(
\begin{array}{ccc}
\cos{\a} & \sin{\a} & 0\\
-\sin{\a} & \cos{\a} & 0 \\
0 & 0 & 1 \end{array} \right)
\left(
\begin{array}{c} 
e_1 \\ e_2 \\ e_3
\end{array} \right) \, .
\eeq
One now finds
\beq
\cn^\prime{}^1{}_{2 \, \sst{(0)}} (x^i) = {\cos{\th} \sin{(2\a)} -\sin{\th}\cos{(2\a)}  \over  1 +\sin{(2\a)} \sin{\th} + \cos{(2 \a)}\cos{\th}} = {\sin{(2\a -\th)} \over 1 + \cos{(2\a -\th)}} \, ,
\eeq
which shows that the new singular surface $\Sigma^\prime$ corresponding to the rotated basis is now located at 
$\th(x^i) - 2 \a(x^i) = +\pi$. This shows that $\S$ is similar to a `Dirac string' singularity: it is a gauge dependent singular submanifold, whose location can be shuffled around by using a gauge transformation. The non--analytic (actually singular) $\cn\rightarrow \infty$, ... behavior of the Iwasawa variables on $\S$ is evidently problematic within a Fuchsian system approach, because it obliges us to work in an open region $U$ of $\RR^3$ which does not contain $\Sigma$. The fact we just showed that the location of $\Sigma$ can be moved around means that we can essentially bypass this technical problem by using simultaneously two separate Fuchsian systems, corresponding to different choices of underlying frames 
$(\o^i,e_i)$ in the analytic spatial manifold $M_d$, yielding finite values Iwasawa variables in two complementary open regions $U$ and $U^\prime$. Such a construction bypasses the analyticity problems near $\Sigma$ and $\Sigma^\prime$. However, the codimension 2 submanifold $\cl$ (where two eigenvalues coincide) remains excluded from these two complementary Fuch analyses. In other words, the present Iwasawa variables--based approach discussed here cannot cover in an analytic manner the measure--zero submanifolds of $M_d$ where 2 (or more generally $n\geq 2$) eigenvalues coincide. We leave this technical problem to further analyses. \newline

In this respect let us remark that the root of the problems linked, either in the AVTD or the Iwasawa approaches, to coinciding eigenvalues of $k_{ij}$ admits a simple geometrical interpretation. The crucial point is that the eigenvectors of $K$, considered as functions of the auxiliary angle $\th$ introduced above ($a= \rho \cos{\th}, \, b = \rho \sin{\th}$) depend on $\th$ in the same manner as a spinor would transform under a $SO(3)$ rotation of angle $\th$. Indeed, the diagonalisation of the matrix $K$ is easily checked to yield the following eigenvectors (with respect to the orthonormal basis $e_1,\, e_2, \, e_3$ in which $K$ is expressed),
\beq
v_1 &=& \cos{{\th \over 2}} e_1 + \sin{{\th \over 2}} e_2  \hspace{1cm} \mathrm{with \, eigenvalue \, } c+\rho = c + \sqrt{a^2 + b^2
}\, , \nn \\
v_2 &=& \sin{{\th \over 2}} e_1 - \cos{{\th \over 2}} e_2 \hspace{1cm} \mathrm{with \, eigenvalue \, } c-\rho = c -\sqrt{a^2 + b^2
} \, , \nn \\
v_3 &=& e_3 \hspace{3.5cm} \mathrm{with \, eigenvalue \, } d \, . 
\eeq
Note the appearance of the half angle $\th/2$ in $v_1 $ and $v_2$. This appearance means that if we follow the evolution of a diagonalising frame $v_1,\, v_2, \, v_3$ along a closed loop in space around $\cl$ (assuming the Jacobian of $a(x)$ and $b(x)$ never vanishes), the eigenvectors $v_1(x^i)$ and $v_2(x^i)$ will, upon their return to the same spatial point $x$ (which corresponds to the same values of $a(x^i)$ and 
$b(x^i)$, but to an angle $\th = 2 \pi$ instead of $\th = 0$), take final values opposite to their initial ones: 
\beq 
v_1(\th = 2\pi )  = -v_1(\th=0), \, v_2(\th = 2\pi )  = -v_2(\th=0),\, v_3(\th = 2\pi )  = v_3(\th=0) \, .
\eeq
This phenomenon explains why, in the Iwasawa approach, the presence of a `line' $\cl$ makes itself felt far away from $\cl$ (\ie on the `singular' half--membrane $\S$): indeed there is a non--trivial \emph{holonomy} of Kasner frames around $\cl$.\footnote{Evidently, one could discuss the richer case where there are several independent codimension 2 manifolds of the type of $\cl$, together with more exceptional submanifolds where more eigenvalues coincide.}

\section{An illustrative Fuchsian toy model}

The aim of this appendix is to see, on a concrete example, the relationship between the structure of the source term and that of a solution of a Fuchsian system. The aim is not to sketch mathematical proofs (for which we refer to \cite{Andersson:2000cv,Damour:2002tc, choquet-bruhat}) but to build some physical intuition. In this perspective, we study a toy model which is the most drastic simplification of the system (\ref{fuchsnonchaotic}) one could consider. We want to show that the unique solution of the system that goes to zero goes to zero like the source and not less fast than the source. To handle this, we use an iterative method. Finally, we also investigate a sligtly more involved toy model that is supposed to mimick the effect of the spatial gradients appearing in the system (\ref{fuchsnonchaotic}). 
 
\subsection{A first toy model \label{toy}}

The first toy model we consider is the following, 
\beq
\dot \b -\pi &=& 0 \, , \nn \\
\dot \pi &=&  e^{-2 w \b } \, . \label{toya}
\eeq
If we consider the asymptotic system to be
\beq
\dot \b_{\sst{[0]}}- \pi_{\sst{[0]}}  &=&0\, , \nn \\
\dot \pi_{\sst{[0]}} &=&0 \, . \label{simpliedsysasymp}
\eeq
The order zero solution is $\pi_{\sst{[0]}} = v$ and $\b_{\sst{[0]}} = v \tau  + \b_\circ$. Rewriting the equations (\ref{toya}) in terms of $\bar \b = \b - \b_{\sst{[0]}}$ and $\pi = \bar \pi - \pi_{\sst{[0]}}$  gives 
\beq
\dot {\bar \b} -\bar \pi&=& 0 \, , \nn \\
\dot {\bar \pi} &=&  e^{-2 w (v \tau + \b_\circ)} e^{-2w(\bar \b)} \, . \label{simpliedsys}
\eeq
The idea is that $\bar \b$ and $\bar \pi$ go to zero as $\tau \rightarrow +\infty$. We can try to solve the system by iteration assuming that $\bar \b$ is small. The first iteration is obtained by replacing $\bar \b $, in the source term of equations (\ref{simpliedsys}),  by its first order estimate which is zero, \ie replacing $e^{-2w(\bar \b)}$ by 1:
\beq
\dot {\bar \b}_{\sst{[1]}} - \bar \pi_{\sst{[1]}}   &=&0\, , \nn \\
\dot {\bar \pi}_{\sst{[1]}}  &=&  e^{-2 w (v \tau + \b_\circ)} \, . \label{simpliedsys1}
\eeq 
The solution of this system is $\bar \pi_{\sst{ [1]}} = {1\over -2 w v} e^{-2w(v \tau + \b_\circ)} $ and $\bar \b_{\sst{ [1]}} = {1\over (2 w v)^2} e^{-2w(v \tau + \b_\circ)}$ (note that we did not add integration constants since we search solutions that go to zero when $\tau \rightarrow + \infty$). The next step consists in replacing $\bar \b$ in the source term by its first order estimate, 
\beq
\dot {\bar \b}_{\sst{[2]}} -\bar \pi_{\sst{[2]}}  &=&0 \, , \nn \\
\dot {\bar \pi}_{\sst{[2]}}  &=&  e^{-2 w (v \tau + \b_\circ)} e^{-2w(\b_{\sst{[1]}})} \sim e^{-2 w (v \tau + \b_\circ)} (1 -2w(\bar \b_{\sst{[1]}} )) \, . \label{simpliedsys2}
\eeq 
The solution of this system is given by $\bar \pi_{\sst{ [2]}} = {1\over -2 w v} e^{-2w(v \tau + \b_\circ)} -{2w \over 2 w(v)^2 (-4w(v))} e^{-4w(\b_{\sst{[0]}})}$ and 
$\bar \b_{\sst{ [2]}} = {1\over (2 w v)^2} e^{-2w(v \tau + \b_\circ)} -{2 w \over 2 w(v)^2 (4w(v))^2 }e^{-4 w(\beta_{\sst{[0]}})} $. Etc. \newline

Therefore, the solution that vanishes goes to zero like the source. We do not lose a $e^{-\epsilon \tau}$ and we conclude that the Fuchs theorem is too strong in this situation. For this very simple toy model, the exact solution of the system (\ref{toya}) can be written as follows, 
\beq
\tau = \int {d\b \over \sqrt{2E - {1 \over w} e^{-2 w \b}} }\, ,
\eeq
where $E$ is a constant of integration and this integral gives explicitly, 
\beq
\tau  = {\b \over \sqrt{2 E}}  + {1 \over w \sqrt{2E}} \ln{(1 + \sqrt{ 1 - { e^{-2 w \b} \over 2 E w}})} + C  \, , \eeq
where $C$ is a constant of integration. 

\subsection{The effect of spatial gradients }

Let us now consider a less primitive toy model, mimicking walls with coefficients depending explicitly on the spatial derivatives of $\b$, 
\beq
\dot \b &=& \pi \, , \nn \\
\dot \pi &=&  \partial_x( e^{-2 w \b }) \, . \label{simplied2sys}
\eeq
As asymptotic system, we take the same as in the previous example (\ref{simpliedsysasymp}). The order zero solution is $\pi_{\sst{[0]}} = v(x)$ and $\b_{\sst{[0]}} = v (x)\tau  + \b_\circ(x)$. Rewriting equations (\ref{simplied2sys}) in terms of $\bar \b = \b - \b_{\sst{[0]}}$ and $\pi = \bar \pi - \pi_{\sst{[0]}}$  gives 
\beq
\dot {\bar \b} &=& \bar \pi \, , \nn \\
\dot {\bar \pi} &=& -2 w(\partial_x v \tau + \partial_x \b_\circ + \partial_x \bar \b)e^{-2 w (v \tau + \b_\circ)} e^{-2w(\bar \b)} \, . \nn
\eeq
The first iteration is again obtained by putting $\bar \b$ is the left hand side to zero, 
\beq 
\bar{\pi}_{\sst{[1]}}  = e^{-2w(\b_{\sst{[0]}})} ({\partial_x \b_\circ \over w(v)} + {2 \partial_x v\over 2 w^2}) + {\partial_x v \over w(v)} \tau e^{-2w(\b_{\sst{[0]}})}\eeq  
$\b_{\sst{[1]}} $ is also of the form $e^{-2w(\b_{\sst{[0]}})}(a + b \tau)$. Higher order iterations will give higher powers of $\tau$ which are multiplied by  increasing powers of the `walls' $e^{-2\b_{\sst{[0]}}}$, see \cite{Buonanno:1998bi} for the structure of an all--order iterative example of such a Fuchsian system. Therefore, we can suspect that in this case, the solution decreases less quickly than the source by a polynomial in $\tau$.

\end{document}